# Advancing global aerosol forecasting with artificial intelligence

Ke Gui[1,#], Xutao Zhang[1,#], Huizheng Che[1,*], Lei Li[1], Yu Zheng[1], Linchang An[2], Yucong Miao[1], Hujia Zhao[3], Oleg Dubovik[4], Brent Holben[5], Jun Wang[6], Pawan Gupta[7,8], Elena S. Lind[9], Carlos Toledano[10], Hong Wang[1], Zhili Wang[1], Yaqiang Wang[11], Xiaomeng Huang[12], Kan Dai[2], Xiangao Xia[13], Xiaofeng Xu[14], Xiaoye Zhang[1,*]

[1] State Key Laboratory of Severe Weather & Key Laboratory of Atmospheric Chemistry of CMA, Chinese Academy of Meteorological Sciences, Beijing, China

[2] National Meteorological Center, CMA, Beijing, China

[3] Institute of Atmospheric Environment, China Meteorological Administration, Shenyang, China

[4] Université de Lille, CNRS, UMR 8518 – Laboratoire d'Optique Atmosphérique, Lille, France

[5] NASA Goddard Space Flight Center, Greenbelt, Maryland, USA

[6] Department of Chemical and Biochemical Engineering, The University of Iowa, Iowa City, IA, USA

[7] STI, Universities Space Research Association (USRA), Huntsville, Alabama, USA

[8] NASA Marshall Space Flight Center, Huntsville, Alabama, USA

[9] Department of Electrical and Computer Engineering, Virginia Tech, Blacksburg, Virginia, USA

[10] Grupo de Optica Atmosférica, Universidad de Valladolid, Paseo Prado de la Magdalena, Valladolid, Spain

[11] Institute of Artificial Intelligence for Meteorological, Chinese Academy of Meteorological Sciences, Beijing, China

[12] Department of Earth System Science, Tsinghua University, Beijing, China

[13] Key Laboratory of Middle Atmosphere and Global Environment Observation, Institute of Atmospheric Physics, Chinese Academy of Sciences, Beijing, China

[14] China Meteorological Administration, Beijing, China

# These authors contributed equally

* Corresponding authors.

E-mail addresses: chehz@cma.gov.cn (H. Che) and xiaoye@cma.gov.cn (X. Zhang)

**Abstract:** Aerosol forecasting is essential for air quality warnings, health risk assessment, and climate change mitigation. However, it is more complex than weather forecasting due to the intricate interactions between aerosol physicochemical processes and atmospheric dynamics, resulting in significant uncertainty and high computational costs. Here, we develop an artificial intelligence-driven global aerosol-meteorology forecasting system (AI-GAMFS), which provides reliable 5-day, 3-hourly forecasts of aerosol optical components and surface concentrations at a 0.5° × 0.625° resolution. AI-GAMFS combines Vision Transformer and U-Net in a backbone network, robustly capturing the complex aerosol-meteorology interactions via global attention and spatiotemporal encoding. Trained on 42 years of advanced aerosol reanalysis data and initialized with GEOS Forward Processing (GEOS-FP) analyses, AI-GAMFS delivers operational 5-day forecasts in one minute. It outperforms the Copernicus Atmosphere Monitoring Service (CAMS) global forecasting system, GEOS-FP forecasts, and several regional dust forecasting systems in forecasting most aerosol variables including aerosol optical depth and dust components. Our results mark a significant step forward in leveraging AI to refine physics-based aerosol forecasting, facilitating more accurate global warnings for aerosol pollution events, such as dust storms and wildfires.

**Main**

Atmospheric aerosols, arising from both anthropogenic and natural sources, play a critical role in the Earth's climate system, affecting radiative forcing, cloud microphysics, and atmospheric chemistry[1,2]. Due to their diverse optical and microphysical properties, combined with complex chemical compositions, aerosols influence weather and climate in varied ways[3,4]. Key components, such as black carbon (BC) and dust, exhibit significant variability in radiative forcing, making aerosols a major source of uncertainty in climate change assessments[5,6]. Additionally, the complex chemical reactivity and wide particle size ranges in aerosols can degrade air quality[7–9], posing health risks, including respiratory, cardiovascular, and neurological diseases[10]. Accurate forecasting of aerosol distributions and compositions is therefore essential for improving air quality management, protecting public health, and mitigating climate change.

However, aerosol forecasting presents significantly greater complexity and cost than weather forecasting due to the need to account for diverse aerosol sources and types, intricate chemical reactions, physical processes, and multiscale interactions with weather systems[11,12]. These complexities result in nonlinear and highly variable processes for aerosol generation, transport, transformation, and deposition, contributing substantially to forecast uncertainty[13]. To enable short- to medium-term aerosol forecasting, traditional physics-based forecasting systems, such as the Copernicus Atmospheric Monitoring Service (CAMS) from the European Centre for Medium-Range Weather Forecasts (ECMWF)[14] and NASA's Global Earth Observing System Forward

Processing (GEOS-FP)[15], couple numerical weather prediction (NWP) models with atmospheric chemical transport models. These systems must simultaneously resolve atmospheric dynamics and compute thousands of aerosol-related chemical reactions and microphysical interactions, further intensifying the already high computational cost of NWP[16]. Recent advances in artificial intelligence (AI) have opened new avenues, leading researchers to explore advanced neural networks as complementary tools for NWP[17–25] and its downstream tasks, such as oceanic variables[26,27]. These neural network models have demonstrated considerable promise in enhancing computational efficiency and accuracy in weather forecasting; however, AI research specifically targeting global aerosol forecasting remains notably underdeveloped. Although recent studies have begun applying deep learning to aerosol forecasting on both global and regional scales[28,29], these efforts largely depend on NWP inputs and are often restricted to single aerosol metrics, such as total aerosol optical depth (AOD). To date, no AI model has achieved the operational integration required to simultaneously forecast global aerosol components alongside meteorological conditions, underscoring a critical gap in the field and emphasizing the urgent need for innovation in aerosol-meteorology forecasting. To address this gap, we propose the AI-driven global aerosol-meteorology forecasting system (AI-GAMFS), designed to rapidly simulate complex aerosol-meteorology interactions across spatial and temporal scales. The overall architecture is illustrated in Fig.1a.

AI-GAMFS is built on a backbone architecture comprising the core modules of Vision Transformer (ViT) and U-Net, enabling global 5-day forecasts at a 50 km resolution and 3-hour temporal intervals. It forecasts AOD, the optical properties and surface concentrations of key aerosol components—including sulfate, dust, black carbon (BC), organic carbon (OC), and sea salt (SS)—as well as surface and upper-level meteorological variables that govern aerosol lifecycle dynamics. By employing a relay forecasting strategy, AI-GAMFS combines four pre-trained models tailored to specific forecast lead times, effectively mitigating the cumulative error in medium-term forecasting caused by high-frequency iterations. Evaluation experiments conducted in 2023, comparing AI-GAMFS with regional dust models, demonstrate its substantial superiority in forecasting key dust parameters, surpassing current dust storm forecasting systems. Moreover, evaluations based on in-situ observational data from hundreds of the Aerosol Robotic Network (AERONET) stations across the global reveal that AI-GAMFS achieves higher accuracy in AOD forecasting than the state-of-the-art CAMS forecast systems. Notably, when driven by GEOS-FP analyses in an operational configuration, operational AI-GAMFS outperforms conventional GEOS-FP forecasts for nearly all aerosol variables, while reducing computational costs by an order of magnitude. These advancements represent a significant leap in utilizing AI to enhance physics-based aerosol forecasting, thereby facilitating more accurate global warnings for aerosol pollution events, such as dust storms and wildfires.

## Results

**AI-GAMFS**

AI-GAMFS is designed to provide global 5-day aerosol-meteorology forecasts with a temporal resolution of 3 hours (01:30, 04:30, 07:30,..., 22:30 UTC) and a spatial resolution of approximately 50 km, covering 56 variables, including 12 aerosol variables and 42 surface and upper-air meteorological parameters. The model consists of three modules (Fig. 1a): (1) cube embedding, which extracts three-dimensional spatiotemporal features from the input feature matrix; (2) the Vision Transformer (ViT)[30], which employs a multi-head self-attention mechanism to process and understand complex relationships between features; and (3) cube unembedding, which reconstructs high-dimensional features back to the original spatial resolution using deconvolution and upsampling techniques. To ensure the accuracy and fidelity of the forecasts, skip connections are incorporated. Working synergistically, these modules accurately forecast the spatial fields of aerosol and meteorology states at the next time step, using the previous time step as input. A detailed description of the model construction process is provided in the Methods section.

AI-GAMFS was trained using the Modern-Era Retrospective Analysis for Research and Applications, version 2 (MERRA-2) atmospheric reanalysis data, which integrates jointly assimilated meteorological and aerosol observational data. By learning from 42 years of MERRA-2 data—including aerosol optical components, surface concentrations, and key meteorological variables that influence aerosol emissions, transport, and deposition—AI-GAMFS aims to learn a general representation of aerosol-meteorology interactions. We trained four base models with forecast lead times of 3, 6, 9, and 12 hours, respectively. Each base model was trained for 80 epochs using the same framework and settings, containing approximately 1.2 billion parameters, and was trained for 10 days on 8 L40 GPUs. To mitigate error accumulation due to long-term iterations in a single model, a temporal aggregation strategy was employed to perform relay forecasting with the four base models (Fig.1b). Once pre-training and relay connection are completed, the final AI-GAMFS model generated 5-day operational forecasts at approximately 39 seconds on a single L40 GPU, using real-time GEOS-FP analysis fields, achieving a speed several orders of magnitude faster than conventional GEOS-FP forecasts.

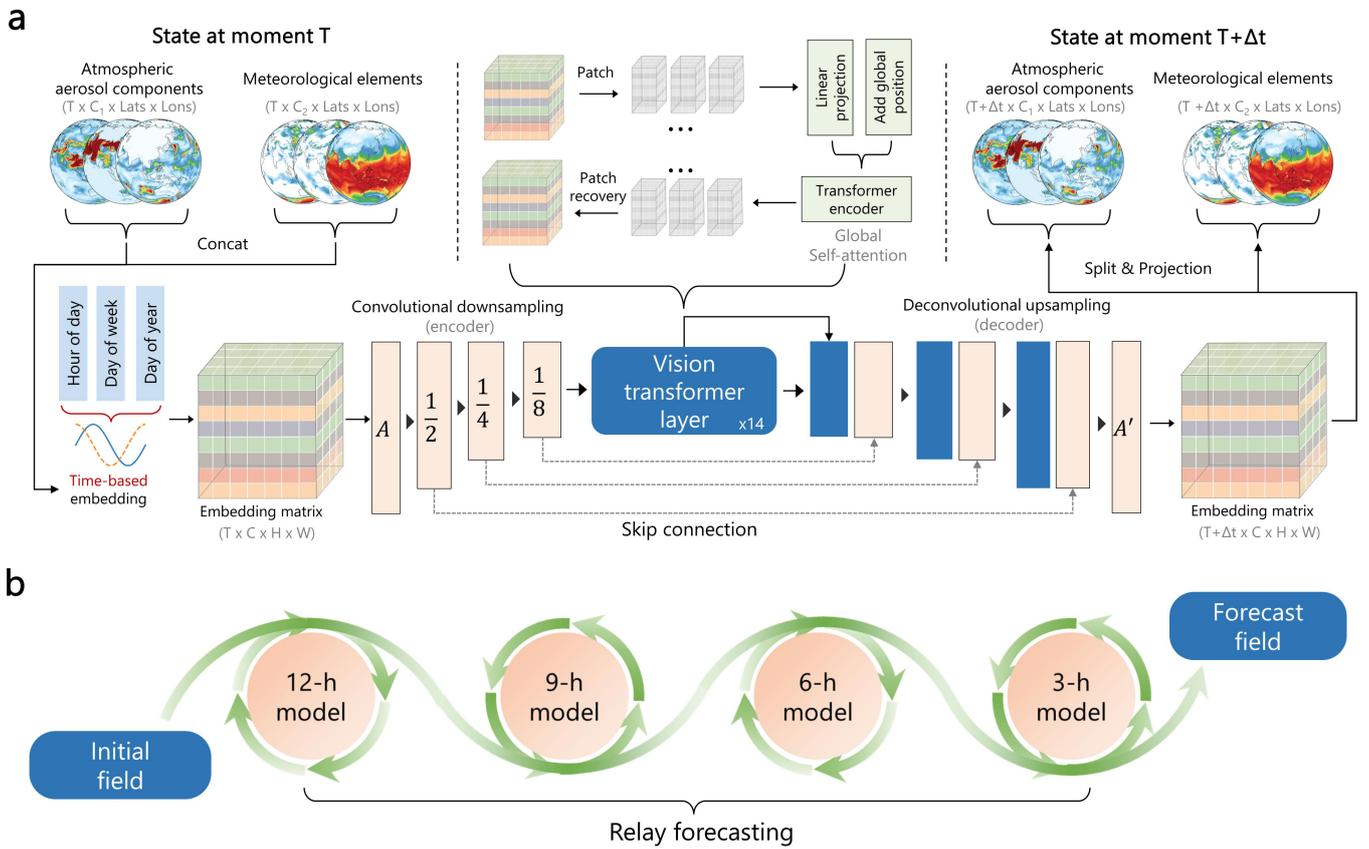

**Fig. 1. Architecture of the AI-driven Global Aerosol-Meteorology Forecasting System (AI-GAMFS). a,** The AI-GAMFS model consists of three primary components: cube embedding, Vision Transformer (ViT), and cube unembedding. **b,** Temporal aggregation strategy used in AI-GAMFS for relay forecasting at specified lead times, achieved by integrating four pre-trained models–3-hour, 6-hour, 9-hour, and 12-hour models–each trained with identical configurations.

**Relay forecasting reduces accumulation errors**

A persistent challenge in current neural network-based atmospheric forecasting is error accumulations as the number of iterations increases, a problem that has been widely recognized in previous studies[19,22]. We adopted a temporal aggregation strategy to perform relay forecasting with pre-trained base models for different forecast lead times (Fig. 1b), aiming to mitigate the error accumulation problem in short- to medium-term forecasts. To achieve this, we trained four base AI-GAMFS models with forecast lead times of 3, 6, 9, and 12 hours, respectively, under identical configurations. To identify the optimal relay forecasting strategy, we designed four progressive forecasting strategies: the 3-hour single-model, the 3- and 6-hour relay, the 3-, 6-, and 9-hour relay, and the 3-, 6-, 9-, and 12-hour relay. Figure 2a illustrates the frequency with which the four pre-trained base models (with lead times of 3, 6, 9, and 12 hours) are invoked under these four forecasting strategies. Notably, for forecasts with a specific lead time, when at least two pre-trained base models are used in the relay, we prioritize models with longer lead times, iteratively using their forecast results as inputs for the next forecast time step, thereby minimizing the number of iterations as much as possible.

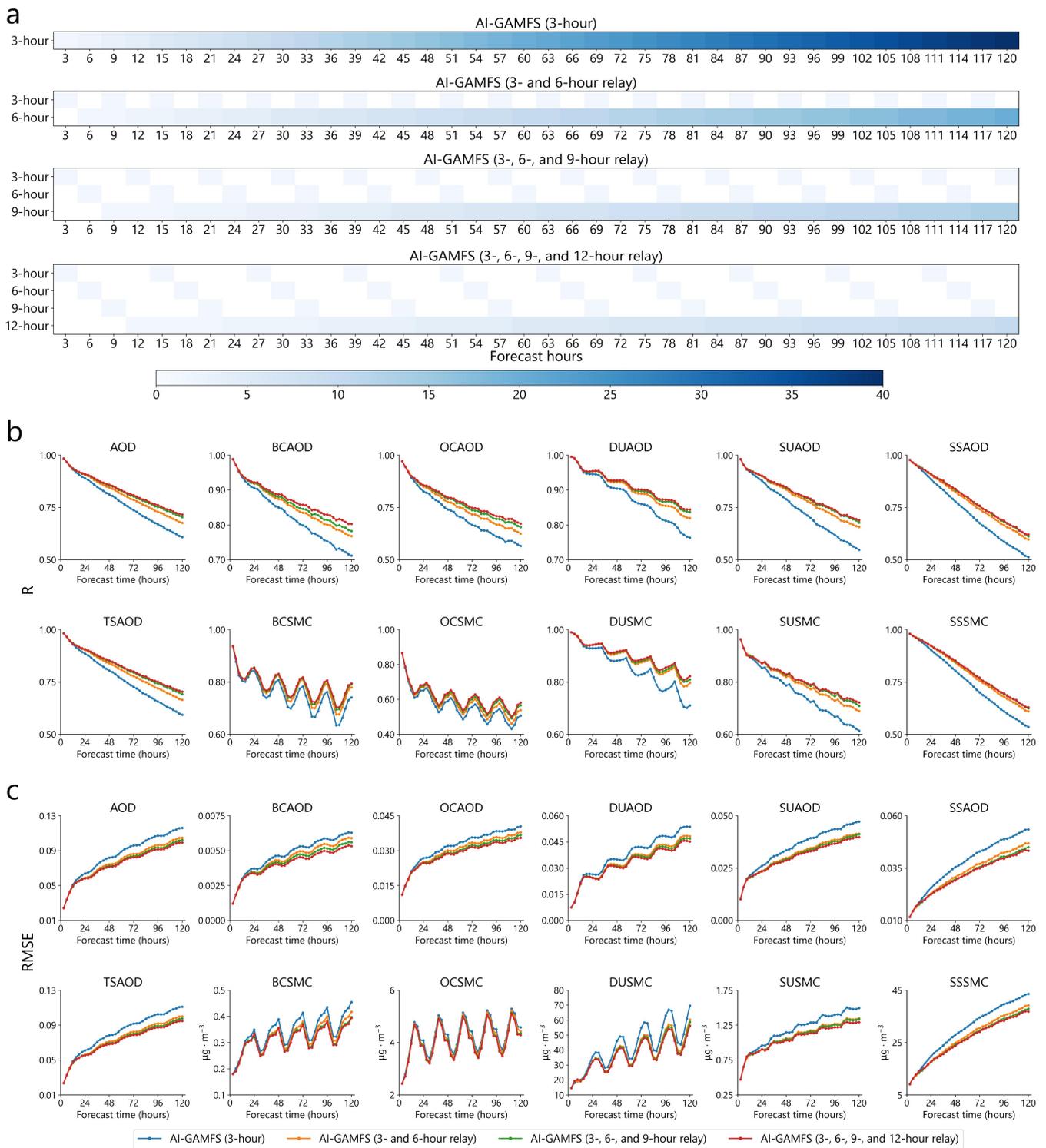

**Fig. 2. Comparison of aerosol forecasting accuracy between single-model and multi-model relay forecasting approaches. a,** The number of times the 3-hour, 6-hour, 9-hour, and 12-hour models are invoked at different forecast lead times, using the 3-hour single model, the 3- and 6-hour relay model, the 3-, 6-, and 9-hour relay model, and the 3-, 6-, 9-, and 12-hour relay model. **b, c,** Accuracy of global 5-day deterministic forecasts for 12 aerosol variables in the 2022 test set, showing spatial $R$ **(a)** and latitude-weighted RMSE **(b)** over time, for the 3-hour single model, the 3- and 6-hour relay model, the 3-, 6-, and 9-hour relay model, and the 3-, 6-, 9-, and 12-hour relay model.

We compared the global 5-day forecasting accuracy of AI-GAMFS, initialized daily at 22:30 UTC, for all 12 aerosol variables using different relay forecasting strategies, with the 2022 MERRA-2 data (test set) as a baseline. The aerosol variables include AOD, total scattering AOD (TSAOD), sulfate AOD (SUAOD), DUAOD, BCAOD, OCAOD, SSAOD, sulfate surface mass concentration (SUSMC), DUSMC, BCSMC, OCSMC, and SSSMC. Figures 2b and 2c show the time series of the global spatial correlation coefficient ($R$) and latitude-weighted root mean square error (RMSE) for these aerosol variables, respectively. The results indicate that, within a 24-hour forecast horizon, the performance of the 3-, 6-, 9-, and 12-hour relay model is similar to that of the 3-hour single-model and other relay models. However, for forecast lead times beyond 24 hours, the 3-, 6-, 9-, and 12-hour relay model show superior accuracy for nearly all aerosol variables, both in terms of $R$ and RMSE. For instance, at a 120-hour lead time, the RMSE values for the 3-, 6-, 9-, and 12-hour relay model are typically 15.1%, 5.6%, and 3.2% lower than the 3-hour single model, the 3- and 6-hour relay, and the 3-, 6-, and 9-hour relay models, respectively. This advantage is also evident in global forecasts for various meteorological variables (Fig. S1). The use of four base models in the relay strategy generally yields results comparable to or slightly better than those from the 3- and 6-hour relay and 3-, 6-, 9-hour relay models, but significantly outperforms the 3-hour single model. However, we note that while the aggregation strategy helps alleviate short- to medium-term error accumulation, the improvement tends to plateau as the number of base models increases. Therefore, we ultimately selected the 3-, 6-, 9-, and 12-hour relay strategy as the final AI-GAMFS model, which was used in all subsequent evaluations and analyses.

**Superior to physics-based regional dust forecasting**

East Asia is one of the regions most severely affected by dust storms worldwide, highlighting the critical need for accurate forecasting. Dust AOD (DUAOD) and dust surface mass concentration (DUSMC) are two key variables used to assess the impact of dust storms on the atmospheric column and surface air quality, and they are commonly used to characterize the intensity of such events. The AI-GAMFS model, which forecasts both DUAOD and DUSMC, offers an opportunity to assess its performance relative to several well-established physics-based dust forecasting models.

For this evaluation, we used East Asia dust forecast products for 2023, derived from forecasts of five physics-based dust forecasting models deployed at the Sand and Dust Storm Warning Advisory and Assessment System (SDS-WAS) Asian regional centre. These models include CAMS, SILAM from the Finnish Meteorological Institute (FMI-SILAM), CUACE/Dust from the China Meteorological Administration (CMA-CUACE/Dust), MASINGAR from the Japan Meteorological Agency (JMA-MASINGAR), and ADAM from the Korea Meteorological Agency (KMA-ADAM). We evaluated the 5-day forecast accuracy of DUAOD and DUSMC from AI-GAMFS (driven by MERRA-2 reanalysis and GEOS-FP analyses, and initialized daily at

22:30 UTC), CAMS, FMI-SILAM, CMA-CUACE/Dust, and KMA-ADAM (initialized daily at 00:00 UTC), relative to MERRA-2 data from 2023. Since JMA-MASINGAR initializes daily at 12:00 UTC and provides 3-day forecasts, we adjusted the initialization time for AI-GAMFS to 10:30 UTC for comparison. Additionally, due to differences in forecast coverage areas and temporal resolutions across models, we conducted the evaluation only for the overlapping East Asia region (see Fig. S2a) and applied time interpolation to the different models (see Methods). Unlike some previous data-driven weather models that are initialized solely with time-lagged reanalysis data, AI-GAMFS uses both MERRA-2 reanalysis data and near-real-time (NRT) GEOS-FP analyses for initialization, with the latter better representing its operational performance in real-world scenarios.

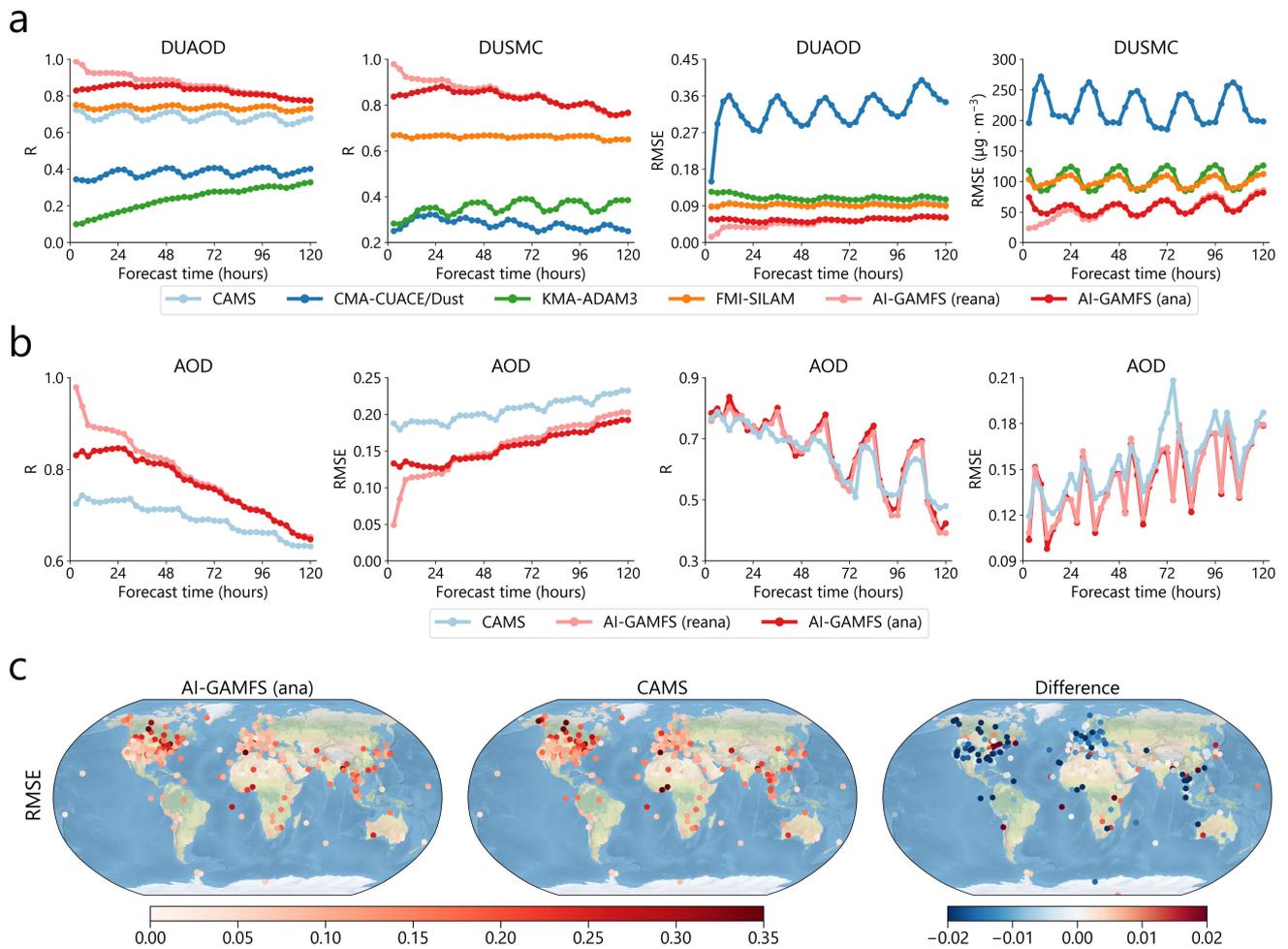

Fig.3. Superior performance of AI-GAMFS in AOD and dust aerosol forecasting compared to CAMS and various regional dust forecasting models. **a,** Using MERRA-2 as the reference baseline, this panel presents a comparison of 5-day deterministic forecast accuracy (evaluated by spatial $R$ and latitude-weighted RMSE) for DUAOD and DUSMC over East Asia in 2023. The comparison includes AI-GAMFS driven by MERRA-2 reanalysis and GEOS-FP analyses, alongside CAMS, CMA-CUACE, KMA-ADAMS, and FMI-SILAM. **b,** A global 5-day deterministic forecast accuracy comparison for AOD in 2023 between AI-GAMFS (driven by MERRA-2 reanalysis and GEOS-FP analyses) and CAMS, using MERRA-2 (left two panels in **b**) and AERONET (right two panels in **b**) as the reference baseline. For AERONET, the evaluation metric for each step is calculated by aggregating all matched samples globally in 2023. **c,** Spatial distribution of the average RMSE for each step of the 5-day AERONET-based AOD forecast (a total of 40 steps) from CAMS and AI-GAMFS (driven by GEOS-FP analyses) in 2023, along with their differences.

Figures 3a and Fig. S2b show the time series of spatial $R$ and latitude-weighted RMSE for these models. Both DUAOD and DUSMC indicate that the reanalysis-driven AI-GAMFS outperforms the analysis-driven AI-GAMFS in dust forecasts over East Asia within 48 hours. However, beyond 48 hours, the forecast performances of both models converges, suggesting that differences in initial conditions have a significant impact on 1–2 day forecasts, but this influence diminishes as the forecast lead time increases. Nevertheless, AI-GAMFS (both reanalysis-driven and analysis-driven) significantly outperforms the five physics-based dust forecast models across all forecast periods at 72 hours (JMA-MASINGAR, Fig. S2b) and 120 hours (other four models, Fig. 2a). Specifically, for the analysis-driven AI-GAMFS, the spatial $R$ for DUAOD at a 72-hour lead time is improved by 11.7%, 17.9%, 34.2%, 106.8% and 202.1% compared to FMI-SILAM, CAMS, JMA-MASINGAR, CMA-CUACE/Dust, and KMA-ADAMS, respectively. At the 120-hour lead time (i.e., 5 days), AI-GAMFS improves by 6.0%, 14.3%, 92.4%, and 135.9% compared to FMI-SILAM, CAMS, CMA-CUACE/Dust, and KMA-ADAMS, respectively. For DUSMC, AI-GAMFS has a latitude-weighted RMSE of 63.2 μg m$^{-3}$ at a 72-hour lead time, which is approximately 38.8%, 46.7%, 70.4%, and 74.7% lower than FMI-SILAM, KMA-ADAMS, CMA-CUACE, and JMA-MASINGAR, respectively. Taking the mega dust storm in northern China in March 2023 as an example, we found that AI-GAMFS can reliably reproduce the entire dust transport process, including the affected areas and intensity (Fig. S3b). This is further confirmed by better statistical metrics compared to other models. More importantly, AI-GAMFS not only forecasts dust transport paths within 1–2 days, but also forecasts enhanced dust emissions in the Gobi Desert up to 5 days in advance. These features are typically challenging to capture in regional dust forecasting models.

**Enhanced performance in global AOD forecasting**

AOD is one of the most widely observed atmospheric aerosol parameters and is extensively used in climate change research, air quality monitoring, and environmental assessments. This study provides a comprehensive evaluation of the 5-day, 3-hourly global AOD forecasts generated by AI-GAMFS, initialized daily at 22:30 UTC, utilizing MERRA-2 evaluation data from 2023. The performance of AI-GAMFS is compared with that of CAMS (initialized at 00:00 UTC), one of the leading global aerosol forecast models, as illustrated in the left two subplots of Fig. 3b. Consistent with dust-related variables, the impact of initial conditions on AI-GAMFS is most pronounced within the first 48 hours, with little to no effect thereafter. Over the 0–120 hour forecast period, the GEOS-FP analysis-driven AI-GAMFS consistently outperforms CAMS in both $R$ and RMSE. Specifically, AI-GAMFS demonstrates a clear advantage during the 0–2 day period, improving $R$ values by 0.10 and reducing RMSE by 0.06 compared to CAMS. However, as the forecast lead time increases, the advantage of AI-GAMFS diminishes. Despite this, at a 120-hour lead time, AI-GAMFS still produces a lower RMSE than CAMS, with a reduction of approximately 17%.

Given the differences in initial conditions between AI-GAMFS and CAMS, AI-GAMFS may benefit from using MERRA-2 as the reference data for evaluation. To ensure a fairer comparison, we additionally used level-2.0 AOD instantaneous observations from AERONET in 2023 (see Methods) to evaluate the 5-day, 3-hourly AOD forecast performance of both AI-GAMFS and CAMS at 297 globally distributed sites. The right two subplots of Fig.3b present the time series of $R$ and RMSE values, calculated from all matched global samples for 2023, at each forecast lead time. Consistent with the evaluation using MERRA-2 as the reference, AI-GAMFS (driven by GEOS-FP analyses) provides more accurate AOD forecasts than CAMS, as evidenced by overall higher $R$ values and lower RMSE values. Statistically, across all 40 forecast steps (3-hour intervals), AI-GAMFS outperforms CAMS at 27 steps for $R$ and 34 steps for RMSE. Figure 3c further illustrates the spatial distribution of the average RMSE for each step of the 5-day AOD forecast (a total of 40 steps) at each AERONET site, comparing CAMS and AI-GAMFS, along with their differences. AI-GAMFS exhibits lower RMSE values than CAMS at 63.6% of the sites, primarily located in the United States, Europe, South America, Africa, and South Asia. These results robustly demonstrate the superior performance of AI-GAMFS in global AOD forecasting.

**Improved operational global aerosol component forecasting**

In addition to forecasting AOD and dust-related properties, AI-GAMFS simultaneously forecasts TSAOD, the optical properties of other aerosol components (i.e., sulfates, BC, OC, and SS), and their surface concentrations. These component forecasts enable precise assessments of their specific impacts on climate, air quality, and public health. We use the operational GEOS-FP model as a reference baseline, as it represents the state-of-the-art in atmospheric aerosol component forecasting and provides output configurations fully consistent with AI-GAMFS. Using MERRA-2 data collected from July to August 2024 as a reference, we evaluate accuracy using spatial $R$ and latitude-weighted RMSE, as shown in Fig. 4. Additional metrics for surface and upper-level meteorological variables are provided in Fig. S4.

The scorecards indicate that AI-GAMFS delivers exceptional forecasting performance across all 12 aerosol variables. For the first 1–3 days, AI-GAMFS outperforms GEOS-FP in all variables and lead times, except for BCSMC and OCSMC at specific time points (based on the $R$-value). At longer lead times, AI-GAMFS consistently outperforms GEOS-FP, except for two SS-related variables: SSAOD and SSSMC. Aerosol component forecasts are highly sensitive to the accuracy of weather forecasts. While AI-GAMFS does not surpass GEOS-FP in the accuracy of certain meteorological variables, such as wind speed, sea level pressure

(SLP), and temperature, improvements in the forecast accuracy of key variables—such as specific humidity and precipitation—which influence aerosol emissions, transformation, and deposition, enable AI-GAMFS to enhance its aerosol simulations (Fig. S4). However, forecast accuracy for wind speed declines beyond 2 days, which negatively impacts the forecast of SS aerosols.

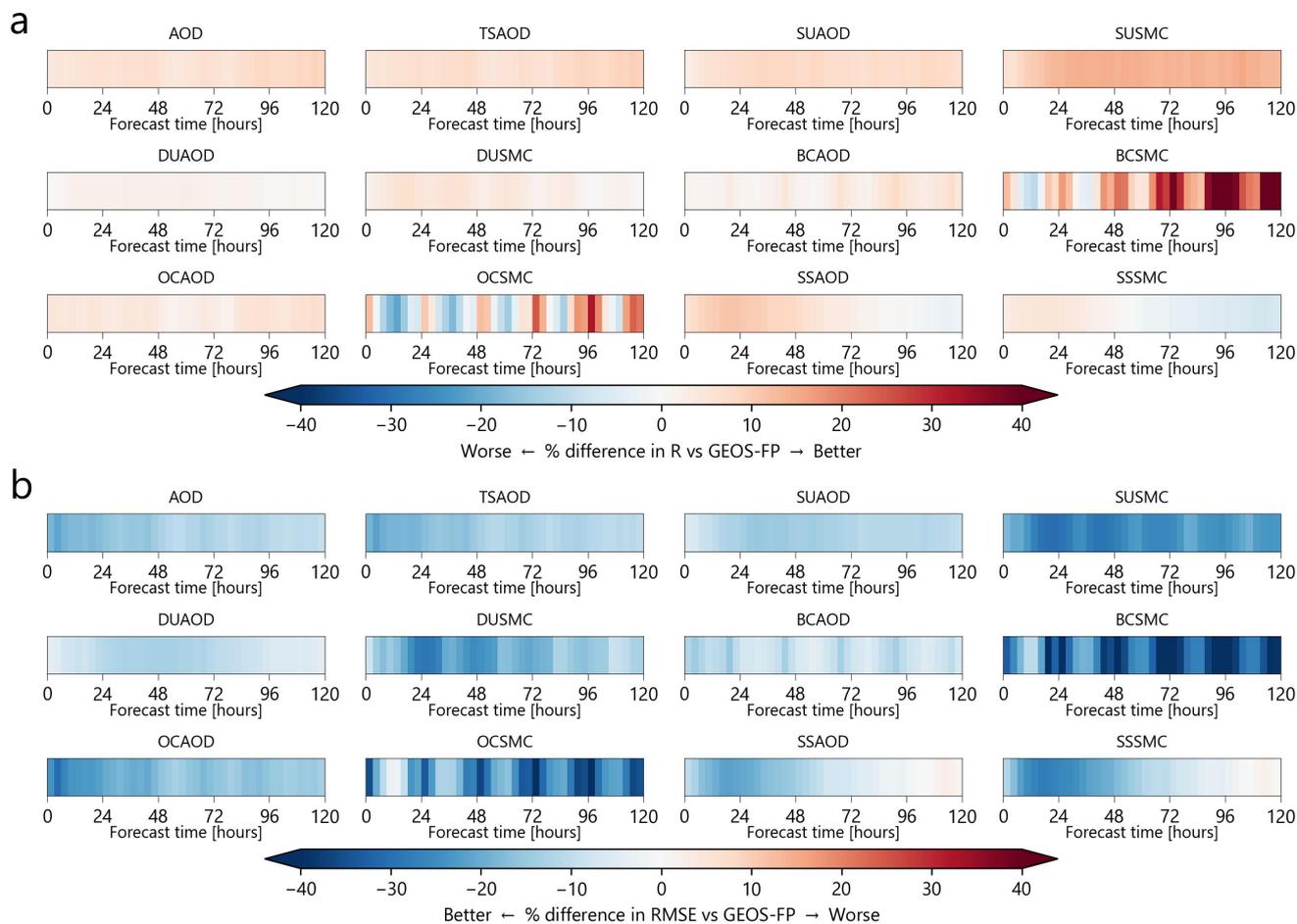

**Fig. 4. AI-GAMFS outperforms operational GEOS-FP in global aerosol component forecasts. a, b,** Scorecards comparing AI-GAMFS against GEOS-FP in 5-day global forecasts of 12 aerosol variables at 3-hour intervals, with spatial *R* **(a)** and latitude-weighted RMSE **(b)** scores from July to August 2024. All AI-GAMFS forecasts are driven by GEOS-FP analyses and evaluated using MERRA-2 reanalysis data from July to August 2024 as the reference baseline.

**Rapid tracking of type-segregated aerosol pollution**

A distinguishing feature of data-driven forecasting is its ability to rapidly track and segregate aerosol pollution types, closely mirroring real-world patterns, at speeds several orders of magnitude faster than traditional physics-based aerosol forecasting models. Figure 5 illustrates a case study with a 3-day lead time, highlighting the performance of AI-GAMFS in forecasting global AOD and its five key components: SUAOD, DUAOD, BCAOD, OCAOD, and SSAOD. A more comprehensive evaluation is provided in Figs. S5–S7, which include spatiotemporal evolution maps for additional aerosol and meteorological variables.

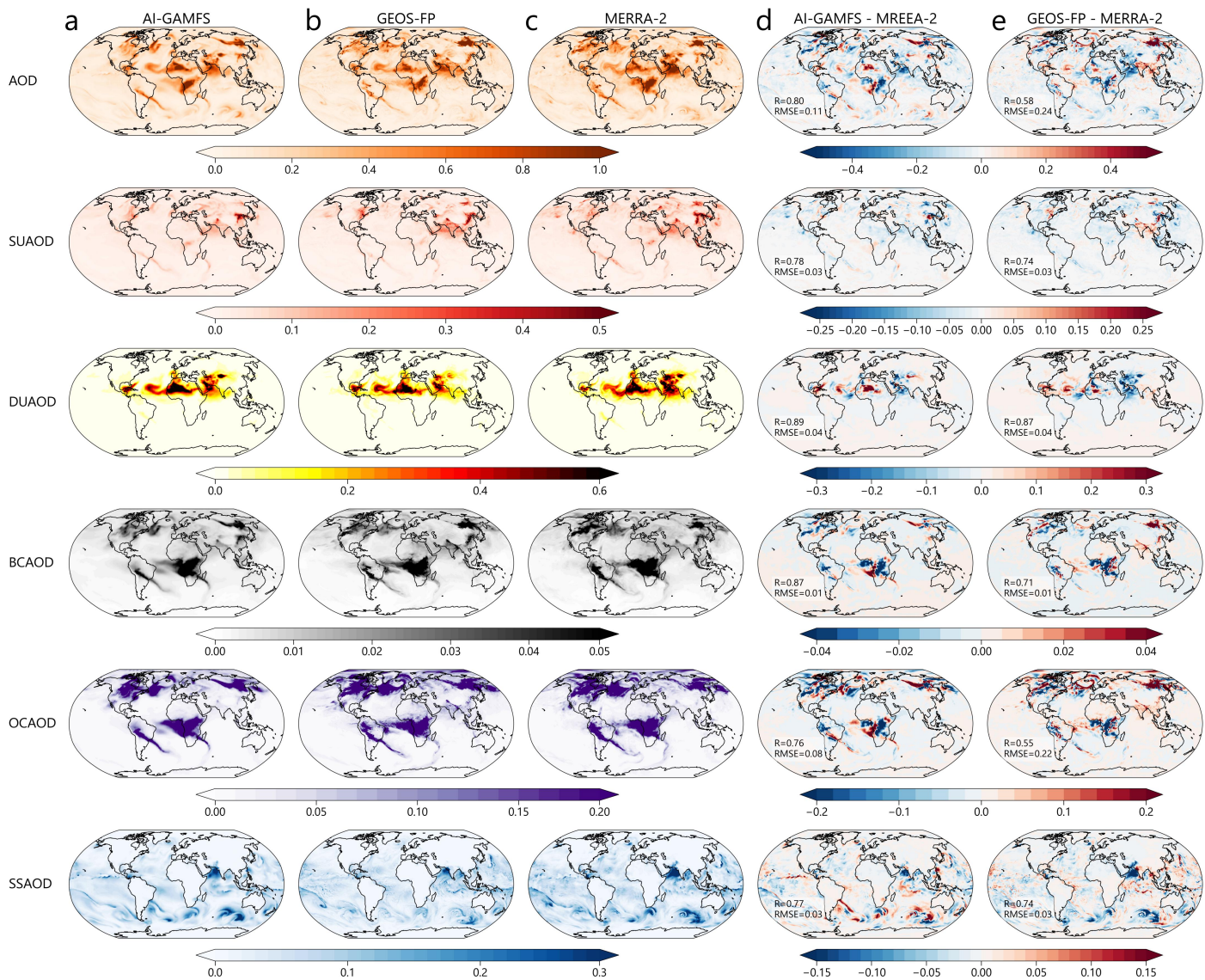

**Fig. 5. Case study of an operational medium-range global aerosol optical component forecasts. a, b, c,** The 3-day lead-time global forecast for AOD and its five key components—SUAOD, DUAOD, BCAOD, OCAOD, and SSAOD—from AI-GAMFS (driven by GEOS-FP analyses) (**a**), GEOS-FP (**b**), and MERRA-2 (**c**), initialized at 22:30 UTC on 26 July 2024. **d, e,** Forecasting errors of AI-GAMFS (**d**) and GEOS-FP (**e**) relative to MERRA-2 reanalysis data. Overall accuracy metrics (i.e., spatial $R$ and latitude-weighted RMSE) for AI-GAMFS and GEOS-FP are also indicated in the lower left corner of each panel in **d** and **e**.

AI-GAMFS produces forecasts that align more closely with MERRA-2 than with GEOS-FP, effectively mitigating spatial smoothing artifacts commonly introduced by longer forecast lead times. In AOD forecasting, AI-GAMFS demonstrates superior $R$ and significantly lower RMSE compared to GEOS-FP. This improvement is consistent across forecasts of various aerosol optical components and surface concentrations (Fig. S5). Additionally, the simultaneous forecasting of aerosol extinction and scattering by AI-GAMFS enables an indirect forecast of the spatiotemporal evolution of aerosol absorption optical depth (AAOD) (Fig. S6), a critical parameter for assessing climate effects[3]. Overall, the exceptional performance of AI-GAMFS is largely attributed to its precise forecasting of key meteorological variables (Fig. S7). Saharan dust and Central African wildfires are long-established sources of global aerosol loading, and their accurate forecasting remains a significant challenge. We further evaluate the performance of AI-GAMFS in forecasting regional dust and BC at one-day interval, as shown in Figs. S8 and S9. Compared to GEOS-FP, AI-GAMFS significantly improves the simulation of Saharan dust and Central African wildfire aerosols, as evidenced by enhanced $R$ and reduced RMSE. Furthermore, AI-GAMFS successfully captures the trans-Pacific transport of dust and smoke aerosols, underscoring its robustness in forecasting long-range aerosol transport. Notably, AI-GAMFS also captures the spatiotemporal evolution of smoke aerosols in South America, in addition to Africa (Fig. 9).

**Discussion**

The development of AI-GAMFS represents a significant advancement in global aerosol forecasting, utilizing AI to extract valuable insights from 42 years of coupled aerosol-meteorology reanalysis data. By learning the complex interactions between aerosols and weather systems, AI-GAMFS demonstrates the potential of AI to propel operational weather forecasting towards more sophisticated environmental meteorological forecasts. We show that AI-GAMFS outperforms several physics-based global and regional aerosol forecasting systems, providing superior deterministic forecasts for key variables such as AOD, DUAOD, and DUSMC. Notably, when initialized with real-time, publicly available GEOS-FP analysis fields, AI-GAMFS delivers more accurate 5-day global AOD forecasts—along with five aerosol optical components and surface concentrations (sulfate, dust, BC, OC, and SS)—in under one minute, compared to GEOS-FP global forecasts. In contrast to physical-based aerosol forecast models, which typically provide no more than two forecasts per day, AI-GAMFS delivers eight forecasts daily, significantly improving forecast timeliness and providing a more accurate representation of the spatiotemporal variations in aerosol components.

Although AI-GAMFS shows substantial potential in refining global aerosol forecasting, there remains considerable room for improvement in model architecture, training strategies, and sample size. For instance, transpose convolution can introduce spatial reconstruction errors during upsampling; incorporating spatial-aware upsampling modules, such as CARAFE (Content-Aware ReAssembly of Features)[31], could enhance the recovery of spatial details. In terms of training strategies, the incorporation of additional prior information typically improves temporal forecasting accuracy. To this end, we intend to adopt a multi-time-step rolling input strategy[22] to further optimize the autoregressive model's performance. Meanwhile, future models should incorporate static or dynamic anthropogenic emission inventories, along with other background information, to mitigate the impact of anthropogenic activities[24]. Finally, the current AI-GAMFS training dataset consists of approximately 120,000 time steps, significantly fewer than other data-driven weather forecasting models, such as Pangu-weather, which uses about 340,000 samples[19]. This limited sample size likely contributes to the model's lower accuracy in forecasting key meteorological variables, including wind speed and temperature, compared to GEOS-FP, thus affecting the accuracy of SS forecasts. To improve overall performance, future work will focus on integrating higher temporal resolution training datasets, where computational resources permit. Additionally, due to limitations in accessing GEOS-FP historical data, we will need to accumulate longer time-series data for cross-seasonal evaluations, enabling a more comprehensive assessment of AI-GAMFS's stability and forecasting capabilities.

The potential implications of AI-GAMFS for the field of atmospheric pollution forecasting are profound. This breakthrough marks a pivotal leap in rapid, high-precision AI-based global aerosol forecasting, with particular significance for economically vulnerable regions and areas frequently affected by dust storms, wildfires, and other forms of air pollution. It has the potential to provide more accurate global aerosol pollution alerts, mitigate public health risks, reduce the economic burden of pollution, and serve as a critical tool for advancing global strategies to combat environmental degradation and climate change.

## Methods

### Datasets details

**MERRA-2 reanalysis.** MERRA-2, developed by NASA's Global Modeling and Assimilation Office (GMAO), is a comprehensive atmospheric reanalysis dataset that spans global atmospheric and climate conditions from 1980 to the present.[32,33]. By assimilating satellite, ground-based, and additional observational data into the Goddard Earth Observing System, version 5 (GEOS-5) Earth system model, MERRA-2 provides high-precision meteorological parameters and multi-layer atmospheric profiles. With its extensive temporal coverage, high spatial resolution (approximately 50 km), and robust consistency, MERRA-2 has become an indispensable tool for climate change research, air quality monitoring, and environmental policymaking. A defining innovation of MERRA-2 is its aerosol dataset, which integrates joint meteorological and aerosol data assimilation (DA). This marks the first time aerosol radiative effects have been directly incorporated into the atmospheric model[32,33], enhancing the fidelity of aerosol-meteorology interactions. MERRA-2 provides high-resolution data across multiple aerosol components—dust, sulfate, black carbon (BC), organic carbon (OC), and sea salt (SS)—with precise parameters such as spatial distribution, optical depth, concentration, and radiative properties.

We use three subsets of the MERRA-2 time-averaged products—aerosol variables (tavg1_2d_aer_Nx), surface atmospheric variables (tavg1_2d_flx_Nx), and upper-air atmospheric variables (tavg3_3d_asm_Nv)—totalling approximately 42 TB to train, test, and evaluate AI-GAMFS, covering 44 years of data from 1980 to 2023. The dataset has a spatial resolution of 0.5° × 0.625° (361 × 576 latitude-longitude grid points). For each subset, only data from timestamps corresponding to the 3-hourly overlapping periods (01:30, 04:30, 07:30,..., 22:30 UTC) are used. We focus on forecasting 12 aerosol variables, including AOD, TSAOD, SUAOD, DUAOD, BCAOD, OCAOD, SSAOD, SUSMC, DUSMC, BCSMC, OCSMC, and SSSMC. All aerosol optical variables are available at a wavelength of 550 nm. Additionally, we forecast 6 surface atmospheric variables and 4 upper-air atmospheric variables at 9 model levels (72, 68, 63, 60, 56, 53, 51, 48, and 45, corresponding to pressure levels of 985, 925, 850, 800, 700, 600, 525, 413, and 288 hPa). Specifically, the 6 surface atmospheric variables are: surface specific humidity (QLML), surface air temperature (TLML), surface eastward wind (ULML), surface northward wind (VLML), SLP, and total precipitation (PRECTOT). The 4 upper-air atmospheric variables are: specific humidity (QV), air temperature (T), eastward wind (U), and northward wind (V). In total, we forecast and evaluate 54 variables. Detailed information on the MERRA-2 variables used in this study is provided in Table S1.

**GEOS-FP analyses and forecasts.** GEOS-FP is a NRT analysis and forecasting system developed by GMAO[15]. This system provides global meteorological and aerosol analyses (i.e., assimilation fields) and generates 5-day (or 10-day) global forecasts, initialized daily at 00:00 UTC (12:00 UTC). It has a grid resolution of approximately 25 km (latitude 0.25°, longitude 0.3125°). GEOS-FP uses the same model configuration as MERRA-2[32], including the simulation of dust, sulfate, BC, OC, and SS via the Goddard Chemistry Aerosol Radiation and Transport (GOCART) model[34,35]. Additionally, GEOS-FP incorporates the assimilation of satellite-based, bias-corrected AOD data[36].

We used three subsets from the GEOS-FP time-averaged analysis and forecast products (see also Table S1), which are consistent with the nomenclature of the MERRA-2 data, to conduct a 5-day comparison experiment between AI-GAMFS historical deterministic and operational forecasts. These subsets contain 54 target variables that fully align with the inputs and outputs of the AI-GAMFS forecasts. To evaluate the forecast performance of AI-GAMFS relative to other global and regional aerosol forecasting models, we used the historical GEOS-FP analyses and MERRA-2 reanalysis data from 22:30 UTC each day in 2023 to drive AI-GAMFS and generate daily 5-day forecasts for the entire year of 2023. In contrast, collecting historical GEOS-FP forecast data is more challenging, as GMAO only archives forecast data for the last two weeks. As a result, we collected only the GEOS-FP analyses at 00:00 UTC and 5-day forecast data (initialized daily at 00:00 UTC) from July to August 2024 for the NRT operational comparison between AI-GAMFS and GEOS-FP. To drive AI-GAMFS and conduct the comparison analysis, we used bilinear interpolation to resample the GEOS-FP analysis and forecast data to match the spatial resolution of 0.5° × 0.625°.

**CAMS aerosol forecasts.** CAMS, developed by ECMWF, is one of the most advanced global aerosol forecasting systems[14]. It provides twice-daily forecasts of global atmospheric composition, including 5-day forecasts of AOD and DUAOD. Using DA techniques, CAMS integrates prior forecasts with current satellite observations to derive optimal initial conditions. It then applies a numerical atmospheric model based on physical and chemical principles to forecast the evolution of aerosol and other atmospheric compositions over the next 5 days[14,37]. The spatial resolution of the CAMS aerosol forecast product at a single level is 0.4° × 0.4°, with a temporal resolution of 1 hour.

In this study, CAMS serves as the baseline for global AOD forecasts based on a physical model, facilitating a comprehensive comparison with AI-GAMFS. We use the 5-day global AOD forecasts for the entire year of 2023, initialized daily at 00:00 UTC. To align with AI-GAMFS for comparison or analysis, we resampled the CAMS forecast data to match a spatial resolution of 0.5° × 0.625° and a temporal resolution of 3 hours, using time interpolation and bilinear interpolation.

**Physical-based dust forecasts.** In this study, we used the 2023 dust forecast products from five physical-based dust forecasting models developed by various institutions and deployed in SDS-WAS Asian regional centre. These products include two global models: CAMS and FMI-SILAM[38], with FMI-SILAM having a temporal resolution of 1 hour and a spatial resolution of 0.2° × 0.2°. Additionally, we analyzed three regional models: CMA-CUACE/Dust[39], with a temporal resolution of 3 hours and a spatial resolution of 0.5° × 0.5°; JMA-MASINGAR[40], with a temporal resolution of 1 hour and a spatial resolution of 0.5° × 0.5°; and KMA-ADAM3[41], with a temporal resolution of 3 hours and a spatial resolution of 0.5° × 0.5° (see Table S1). The JMA-MASINGAR model provides forecasts with a 3-day lead time, while all other models provide forecasts with a 5-day lead time. Detailed descriptions of these models can be found in their respective technical documentation[14,38–41].

Due to differences in initialization times and dust output variables across the models, we used DUAOD forecast outputs from CAMS, FMI-SILAM, CMA-CUACE/Dust, JMA-MASINGAR, and KMA-ADAM3, while for DUSMC, we utilized outputs from all models except CAMS. Notably, except for JMA-MASINGAR, which initializes at 12:00 UTC, all other forecast products begin at 00:00 UTC. To facilitate comparison, we unified the spatial and temporal resolutions of all model outputs to match the AI-GAMFS resolution of 0.5° × 0.625° spatially and 3 hours temporally.

**AERONET measurement.** AERONET is a global aerosol observation network that provides high-quality ground-based measurements of aerosol optical properties[42]. The network consists of numerous automated stations equipped with sun photometers to monitor AOD and other aerosol parameters in real time. AERONET data are widely regarded as the "gold standard" in atmospheric aerosol observations, serving as high-precision references for climate studies, air quality monitoring, and satellite remote sensing validation. In this study, we used instantaneous AOD observation data (version 3.0, level 2.0)[43] from 297 AERONET sites worldwide in

2023. To ensure the accuracy of the evaluation, we averaged the AERONET instantaneous observations within a half-hour window before and after the forecast lead time, which served as the reference truth. Since AERONET does not provide AOD measurements at 550 nm, we used the following quadratic polynomial interpolation method[44,45] to convert AOD observations at four adjacent wavelengths (440, 500, 675, and 870 nm) into AOD values at 550 nm.

$$\ln_{(\tau_\lambda)} = a_0 + a_1 \ln(\tau_\lambda) + a_2 [\ln(\tau_\lambda)]^2 \quad (1)$$

where $a_0, a_1,$ and $a_2$ represent the fitting coefficients, and $\tau_\lambda$ denotes the AOD values at the respective wavelengths.

**AI-GAMFS architecture**

As illustrated in Fig. 1a, the AI-GAMFS architecture consists of three primary modules: cube embedding, Vision Transformer (ViT), and cube unembedding. The base model of AI-GAMFS is an autoregressive model that uses the spatial feature tensor at the previous time step ($X_{t-n}$) as input to forecast the spatial feature tensor at the next time step ($X_t$). Here, $t-n$ and $t$ represent the previous and upcoming time steps, respectively. The base model considers time steps of 3, 6, 9, and 12 hours. Using the output of the base model as input, AI-GAMFS can generate forecasts for different lead times. Below is a detailed description of the modeling process for each of the aforementioned modules.

**Cube embedding module.** The input spatial field is represented as $X_{t-n}^{54 \times 361 \times 576}$, where 54, 361, and 576 correspond to the total number of input variables (42 meteorological and 12 aerosol variables), the latitude grid points, and the longitude grid points, respectively. In addition to the fixed MERRA-2 features, we also incorporate temporal features such as the hour, the day of the week, and the day of the year. These time features undergo sinusoidal and cosine transformations to better represent the periodic temporal characteristics[24]:

$$sin_{\text{hour}} = \sin\left(2\pi \frac{\text{hour}}{24}\right), sin_{\text{day of week}} = \sin\left(2\pi \frac{\text{day of week}}{7}\right), sin_{\text{day of year}} = \sin\left(2\pi \frac{\text{day of year}}{365.25}\right) \quad (2)$$

$$cos_{\text{hour}} = \cos\left(2\pi \frac{\text{hour}}{24}\right), cos_{\text{day of week}} = \cos\left(2\pi \frac{\text{day of week}}{7}\right), cos_{\text{day of year}} = \cos\left(2\pi \frac{\text{day of year}}{365.25}\right) \quad (3)$$

The processed time-encoded features, serving as dynamic field features, are superimposed on the spatial field features. At each grid point, the corresponding time feature is added, resulting in a spatiotemporal encoded feature matrix of $X_{t-n}^{60 \times 361 \times 576}$. To address the issue of edge effects on Earth, we applied spatial padding to the multidimensional feature matrix, including the time features, padding it to a uniform size of $X_{t-n}^{60 \times 576 \times 576}$. The padded matrix was then input into an initial convolution layer to extract the first set of features, resulting in a standardized tensor $X_{t-n}^{640 \times 288 \times 288}$. This tensor was subsequently fed into a three-layer composite convolutional network for feature upsampling. Each composite convolutional layer consisted of three convolution operations:

one with a 1×1 convolution kernel, one with a 3×3 convolution kernel, and another with a 1×1 convolution kernel. After feature extraction in each layer, the spatial dimensions were downsampled by a factor of 2, and the feature dimension was doubled. After 3 layers, the feature matrix size become $X_{t-n}^{8\times 640\times 36\times 36}$.

**ViT module.** Compared to traditional convolutional neural networks, the ViT can more effectively capture global contextual relationships through self-attention mechanisms, thereby enhancing the understanding of complex input feature patterns[30]. For ease of explanation, we denote the feature matrix extracted by the composite convolutional network as $X^{C\times H\times W}$. $C$ represents the number of channels encoded by the cube embedding module, which is $8\times 640$. $H$ and $W$ are the encoded height and width, both of which are 36. In the ViT module, we segment the input feature tensor into fixed-size patches (each 2×2), resulting in $\frac{HW}{P^2}$ patches. Each patch is flattened into a vector and linearly transformed into a fixed-dimensional space, yielding the feature representation for each patch:

$$X_p = \text{Flatten}(X_p)\cdot W_e \quad (4)$$

where $X_p$ is the $p$-th patch and $W_e$ is the linear transformation matrix. Since the Transformer is sensitive to the input order and we aim for the model to better account for global positional information, we add positional encoding vectors to the mapped patches:

$$z = [X_p^1; X_p^2; ...; X_p^N] + E_{pos} \quad (5)$$

The ViT module calculates the relationship between patches using global multi-head self-attention (MHSA) mechanism[46]. A "head" refers to a set of parallel self-attention computation units. For each attention $\text{head}_i$, we construct Query ($Q$), Key ($K$), and Value ($V$) matrices:

$$Q_h = zW_Q^h, K_h = zW_k^h, V_h = zW_v^h \quad (6)$$

The self-attention scores for each head are then computed (after normalization) as follows:

$$\text{Attention}(Q, K, V) = \text{softmax}\left(\frac{QK^T}{\sqrt{d_k}}\right)V \quad (7)$$

Where $d_k$ is the dimension of the key vectors. The outputs from all attention heads are then concatenated and linearly transformed:

$$\text{MHSA}(z) = \text{Concat}(\text{head}_1, \text{head}_2, ..., \text{head}_h)W_o \quad (8)$$

Here, $W_o$ is the number of attention heads and has a shape of $(h\cdot d_v)\times d_{c'}$, where $h$ is the number of attention heads, $d_v$ is the input feature dimension, and $d_{c'}$ is the dimension of each attention head. In AI-GAMFS, each Transformer encoder layer consists of self-attention, feed-forward networks, layer normalization, and residual connections. We omit the class token typically used in the traditional ViT framework, utilizing it solely for deep

feature extraction:

$$z' = \text{LN}(z + \text{MHSA}(z)), \quad z'' = \text{LN}(z' + \text{FFN}(z')) \quad (9)$$

where, LN denotes layer normalization and FFN is the feed-forward network. Here, $z$ represents the features input to the ViT encoding layer, $z'$ represents the features after passing through the MHSA mechanism and LN, and $z''$ represents the features after further processing through the FFN and LN based on $z'$.

The ViT structure in AI-GAMFS stacks 14 layers, each with 16 attention heads. After the ViT deep feature transformation module, patch recovery reshapes the output back to the original size of the feature matrix extracted by the composite convolution.

**Cube unembedding module.** We then perform four transposed convolutional upsamplings operations on the feature matrix, each time doubling the spatial dimensions and halving the channel dimensions. During the upsampling process, we adopt a U-Net structure, utilizing skip connections at $l$ layer to capture larger-scale composite features:

$$X_l = \text{Concat}(X_l, X_{l'}^{\text{skip}}) \quad (10)$$

At the third layer's output, the feature matrix size is $X_t^{320 \times 288 \times 288}$. Corresponding to the initial convolution in the encoder, we also construct an output convolution at the end of the unembedding module. This removes the padding introduced during convolution, yielding the spatial fields for aerosol and meteorology states, which are used to predict the next time step.

**Training strategy**

We utilize the MERRA-2 reanalysis with a 3-hour temporal resolution to train the AI-GAMFS model. Data from 1980 to 2021 were used for training, data from 2022 served as the test set, and data from 2023 were used for validation. The model employs a rolling training approach, where pairs of samples from two consecutive time points ($X_{t-n}$ and $X_t$) are iteratively fed into the model for training.

For the standardized samples, the mean absolute error (MAE) was used as the loss function:

$$L_1 = \frac{1}{C \times H \times W} \sum_{c=1}^{C} \sum_{i=1}^{H} \sum_{j=1}^{W} |\widehat{X}_{c,i,j}^t - X_{c,i,j}^t| \quad (11)$$

Where $C, H,$ and $W$ denote the number of variables, the latitude grid points, and the longitude grid points, respectively. $X_{c,i,j}^t$ and $\widehat{X}_{c,i,j}^t$ represent the "ground truth" (i.e., MERRA-2) and the forecasted value at the specified forecasting time, respectively.

The AI-GAMFS framework was implemented on the PyTorch platform. Each model, corresponding to a specified lead time and containing approximately 1.2 billion parameters, was trained on a server equipped with

8 L40 GPUs for 80 epochs (approximately 10 days). We used the Adam optimizer with $\beta_1 = 0.9$ and $\beta_2 = 0.999$, an initial learning rate of $3\times10^{-4}$, which was decayed using a cosine annealing schedule to 0.0001 of its initial value. Training was conducted in 32-bit floating-point precision with a dropout rate of 0.15 to mitigate overfitting.

**Forecasting strategy**

Similar to physics-based forecasting models, we observed that forecast errors in deep learning models accumulate and amplify as the number of rolling iterations increases. To mitigate this, we adopted the temporal aggregation strategy from Pangu-Weather[19], which reduces the number of model iterations without compromising the forecast time resolution. Using the same modeling framework and configurations, we train four pre-trained AI-GAMFS models with lead times of 3, 6, 9, and 12 hours, referred to as the 3-hour, 6-hour, 9-hour, and 12-hour models, respectively. For forecasts at specific intervals, we prioritize the long timescale models and combine them with short timescale models in a relay fashion (Fig.1b and Fig. 2a). Unlike traditional purely rolling iterative forecasts, our relay forecasting restarts from several fixed previous time points rather than directly from the preceding time step. We denote the model as $F$, and this relay forecasting strategy can be expressed as:

$$X_t = F_3^{(n_3)}(F_6^{(n_6)}(F_9^{(n_9)}(F_{12}^{(n_{12})}(X_{t_0})))) \quad (12)$$

where $X_{t_0}$ is the initial time and $X_t$ is the specified forecasting lead times. $n_{12}$ is the integer quotient of the forecast duration ($X_t$-$X_{t_0}$) divided by 12, $n_9$ is the integer quotient of dividing the remainder when $X_t$-$X_{t_0}$ is divided by 12 by 9, $n_6$ is the integer quotient of dividing the remainder when $X_t$-$X_{t_0}$ is divided by 9 by 6, and $n_3$ can be either 0 or 1. Although this strategy sacrifices some computational efficiency, it takes advantage of the high-speed capabilities of GPUs, enabling the model to produce a 5-day forecast in approximately 39 seconds on a single L40 GPU.

**Evaluation experiment**

To rigorously evaluate the forecasting capabilities of AI-GAMFS, we conducted a series of evaluation experiments, using MERRA-2 reanalysis and AERONET data as reference baselines.

**AI-GAMFS relay forecast evaluation.** We compared four AI-GAMFS model configurations on the 2022 test set, encompassing all 54 aerosol and meteorological variables. These configurations included: a 3-hour single model, a 3- and 6-hour relay model, a 3-, 6-, and 9-hour relay model, and a 3-, 6-, 9-, and 12-hour relay model. This evaluation provides insight into the optimal relay configurations for enhanced predictive performance.

**AI-GAMFS vs. regional dust forecasting models.** We evaluated AI-GAMFS forecasts against five physics-based dust forecasting models across East Asia, using the 2023 validation dataset. The models included in this comparison—CAMS, CMA-CUACE/Dust, FMI-SILAM, JMA-MASINGAR, and KMA-ADAM3—are either specialized dust forecasting models or aerosol models with dust-specific outputs. The evaluation focused on two critical parameters: DUAOD and DUSMC, which allowed us to evaluate AI-GAMFS's accuracy and reliability in forecasting dust storm events.

**AI-GAMFS vs. CAMS in global AOD forecasts.** In 2023, we conducted a spatial comparison of AI-GAMFS and CAMS (one of the world's state-of-the-art aerosol forecasting models) in their 5-day AOD forecasts at both global and regional scales. Additionally, the 5-day AOD forecasts from AI-GAMFS and CAMS were further evaluated against AERONET observations collected worldwide throughout 2023.

**Operational performance of AI-GAMFS vs. GEOS-FP.**

AI-GAMFS is designed for real-time operational forecasting and utilizes GEOS-FP real-time analyses to generate global 5-day aerosol-meteorology forecasts. To evaluate its operational forecasting capabilities for various aerosol components and meteological variables, we analyzed GEOS-FP forecast outputs for July and August 2024. A detailed comparative assessment of AI-GAMFS and GEOS-FP was performed using MERRA-2 as reference baseline, focusing on all 54 target aerosol and meteorological variables.

**Evaluation metrics.** For the site-scale evaluation, using AERONET as the reference baseline, we employ two metrics: simple RMSE and Pearson's $R$. For the spatial evaluation, using MERRA-2 as the reference baseline, we use two metrics: latitude-weighted RMSE and spatial $R$, defined as follows:

$$\text{RMSE}(c, t) = \sqrt{\frac{\sum_{i=1}^{N_{lat}} \sum_{j=1}^{N_{lon}} w_i (\widehat{X}_{i,j,t} - X_{i,j,t})^2}{N_{lat} \times N_{lon}}} \quad (13)$$

$$R(c, t) = \frac{\sum_{i=1}^{N_{lat}} \sum_{j=1}^{N_{lon}} (\widehat{X}_{i,j,t} - \overline{X}_{i,j,t})(X_{i,j,t} - \overline{X}_{i,j,t})}{\sqrt{\sum_{i=1}^{N_{lat}} \sum_{j=1}^{N_{lon}} (\widehat{X}_{i,j,t} - \overline{X}_{i,j,t})^2} \times \sqrt{\sum_{i=1}^{N_{lat}} \sum_{j=1}^{N_{lon}} (X_{i,j,t} - \overline{X}_{i,j,t})^2}} \quad (14)$$

Where $w_i = N_{lat} \times \frac{cos\phi_i}{\sum_{i=1}^{N_{lat}} cos\phi_i}$, c represents the specified variable, $w_i$ denotes the latitude weight, and $\phi_i$ refers to the latitude value. $\widehat{X}_{i,j,t}$ and $X_{i,j,t}$ correspond to the forecast and "ground truth" values, respectively, for a grid point at a given time.

**Data availability**

All training and validation data supporting the development of AI-GAMFS are publicly available. MERRA-2 reanalysis are available at https://disc.gsfc.nasa.gov/. GEOS-FP aerosol and meteorological analyses, as well as 5-day forecasts initialized daily at 00:00 UTC, are available from

https://portal.nccs.nasa.gov/datashare/gmao/geos-fp/. The CAMS global aerosol forecasts initialized at 00:00 UTC are available from the Copernicus Atmosphere Data Store (ADS) (https://ads.atmosphere.copernicus.eu/datasets/cams-global-atmospheric-composition-forecasts?tab=overview). Subject to access permission, regional dust forecasting data in 2023 from CAMS, CMA-CUACE/Dust, FMI-SILAM, JMA-MASINGAR, and KMA-ADAM3 are available from Sand and dust storm warning advisory and assessment system (SDS-WAS) asian regional center (http://www.asdf-bj.net/). The 2023 global instantaneous AOD observational data (version 3.0, level 2.0) from AERONET sites are available at https://aeronet.gsfc.nasa.gov/.

## Acknowledgements

We thank NASA for providing the MERRA-2 reanalysis, GEOS-FP analysis and forecast data, and AERONET observational datasets. Our sincere gratitude also goes to ECMWF for supplying the CAMS aerosol forecast products and to SDS-WAS for the regional dust forecast products. This research was supported by the National Natural Science Foundation of China (42090033, 42030608, 42175153, 42375188, 42105138, and 42275195), the National Key Research and Development Program of China (2023YFC3706305), the Youth Innovation Team of China Meteorological Administration (CMA2024QN13), and the Basic Research Fund of CAMS (2023Z021).

## Competing interests

The authors declare that they have no competing interests.

## References

1. Prospero, J. M. *et al.* The atmospheric aerosol system: An overview. *Rev. Geophys.* **21**, 1607–1629 (1983).
2. Charlson, R. J. *et al.* Climate Forcing by Anthropogenic Aerosols. *Science.* **255**, 423–430 (1992).
3. Li, J. *et al.* Scattering and absorbing aerosols in the climate system. *Nat. Rev. Earth Environ.* **3**, 363–379 (2022).
4. Bellouin, N. *et al.* Bounding Global Aerosol Radiative Forcing of Climate Change. *Rev. Geophys.* **58**, 1–45 (2020).
5. Zhang, X. *et al.* Aerosol components derived from global AERONET measurements by GRASP: A new value-added aerosol component global dataset and its application. *Bull. Am. Meteorol. Soc.* (2024).
6. Dubovik, O. *et al.* Variability of absorption and optical properties of key aerosol types observed in worldwide locations. *J. Atmos. Sci.* **59**, 590–608 (2002).
7. Che, H. *et al.* Aerosol optical and radiative properties and their environmental effects in China: A review.

*Earth-Science Rev.* **248**, (2024).

8. Menon, S. *et al.* Aerosol climate effects and air quality impacts from 1980 to 2030. *Environ. Res. Lett.* **3**, (2008).

9. Gui, K. *et al.* The Significant Contribution of Small-Sized and Spherical Aerosol Particles to the Decreasing Trend in Total Aerosol Optical Depth over Land from 2003 to 2018. *Engineering* **16**, 82–92 (2022).

10. Pöschl, U. Atmospheric aerosols: Composition, transformation, climate and health effects. *Angew. Chemie - Int. Ed.* **44**, 7520–7540 (2005).

11. Morcrette, J. J. *et al.* Aerosol analysis and forecast in the european centre for medium-range weather forecasts integrated forecast system: Forward modeling. *J. Geophys. Res. Atmos.* **114**, (2009).

12. Kukkonen, J. *et al.* A review of operational, regional-scale, chemical weather forecasting models in Europe. *Atmos. Chem. Phys.* **12**, 1–87 (2012).

13. Baklanov, A. & Zhang, Y. Advances in air quality modeling and forecasting. *Glob. Transitions* **2**, 261–270 (2020).

14. Peuch, V. H. *et al.* The Copernicus Atmosphere Monitoring Service from Research to Operations. *Bull. Am. Meteorol. Soc.* **103**, E2650–E2668 (2022).

15. Lucchesi, R. File Specification for GEOS FP (Forward Processing). **4**, (2018).

16. Hendricks, J., Righi, M. & Aquila, V. Global Atmospheric Aerosol Modeling. 561–576 (2012). doi:10.1007/978-3-642-30183-4_34

17. Pathak, J. *et al.* FourCastNet: A global data-driven high-resolution weather model using adaptive fourier neural operators. *arXiv preprint arXiv:2202.11214* (2022).

18. Lam, R. *et al.* Learning skillful medium-range global weather forecasting. *Science.* **382**, 1416–1421 (2023).

19. Bi, K. *et al.* Accurate medium-range global weather forecasting with 3D neural networks. *Nature* **619**, 533–538 (2023).

20. Nguyen, T., Brandstetter, J., Kapoor, A., Gupta, J. K. & Grover, A. ClimaX: A foundation model for weather and climate. in *Proceedings of Machine Learning Research* **202**, 25904–25938 (2023).

21. Luo, J., Apr, A. I. & Chen, X. Fengwu: Pushing the skillful global medium-range weather forecast beyond 10 days lead. *arXiv preprint arXiv: 2304.02948* (2023).

22. Chen, L. *et al.* FuXi: a cascade machine learning forecasting system for 15-day global weather forecast. *npj Clim. Atmos. Sci.* **6**, 1–11 (2023).

23. Kochkov, D. *et al.* Neural general circulation models for weather and climate. *Nature* **632**, 1060–1066

(2024).

24. Bodnar, C. *et al.* Aurora: A Foundation Model of the Atmosphere. *arXiv preprint arXiv: 2405.13063* (2024).

25. Lang, S. *et al.* AIFS -- ECMWF's data-driven forecasting system. *arXiv preprint arXiv: 2406.01465* (2024).

26. Xiong, W. *et al.* AI-GOMS: Large AI-Driven Global Ocean Modeling System. *arXiv preprint arXiv: 2308.03152* (2023).

27. Wang, X. *et al.* XiHe: A Data-Driven Model for Global Ocean Eddy-Resolving Forecasting. *arXiv preprint arXiv: 2402.02995* (2024).

28. Nowak, T. E., Augousti, A. T., Simmons, B. I. & Siegert, S. DustNet: skillful neural network predictions of Saharan dust. *arXiv preprint arXiv: 2406.11754* (2024).

29. Cai, A. S., Fang, F., Peuch, V. & Alexe, M. Advancing operational $PM_{2.5}$ forecasting with dual deep neural networks (D-DNet). *arXiv preprint arXiv: 2406.19154* (2024).

30. Dosovitskiy, A. *et al.* An Image Is Worth 16X16 Words: Transformers for Image Recognition At Scale. *ICLR 2021 - 9th Int. Conf. Learn. Represent.* (2021).

31. Wang, J. *et al.* CARAFE: Content-aware reassembly of features. *Proc. IEEE Int. Conf. Comput. Vis.*, 3007–3016 (2019).

32. Buchard, V. *et al.* The MERRA-2 Aerosol Reanalysis, 1980 Onward. Part I: System Description and Data Assimilation Evaluation. *J. Clim.* **30**, 6851–6872 (2017).

33. Buchard, V. *et al.* The MERRA-2 aerosol reanalysis, 1980 onward. Part II: Evaluation and case studies. *J. Clim.* **30**, 6851–6872 (2017).

34. Chin, M. *et al.* Tropospheric aerosol optical thickness from the GOCART model and comparisons with satellite and sun photometer measurements. *J. Atmos. Sci.* **59**, 461–483 (2002).

35. Colarco, P., Da Silva, A., Chin, M. & Diehl, T. Online simulations of global aerosol distributions in the NASA GEOS-4 model and comparisons to satellite and ground-based aerosol optical depth. *J. Geophys. Res. Atmos.* **115**, (2010).

36. Albayrak, A., Wei, J., Petrenko, M., Lynnes, C. & Levy, R. C. Global bias adjustment for MODIS aerosol optical thickness using neural network. *J. Appl. Remote Sens.* **7**, 073514 (2013).

37. Inness, A. *et al.* The CAMS reanalysis of atmospheric composition. *Atmos. Chem. Phys.* **19**, 3515–3556 (2019).

38. Sofiev, M. *et al.* Construction of the SILAM Eulerian atmospheric dispersion model based on the advection algorithm of Michael Galperin. *Geosci. Model Dev.* **8**, 3497–3522 (2015).

39. Gong, S. L. & Zhang, X. Y. CUACE/Dust – An integrated system of observation and modeling systems for operational dust forecasting in Asia. *Atmos. Chem. Phys.* **8**, 2333–2340 (2008).

40. Yukimoto, S. *et al.* A new global climate model of the Meteorological Research Institute: MRI-CGCM3: -Model description and basic performance-. *J. Meteorol. Soc. Japan* **90**, 23–64 (2012).

41. Kim, M., Cho, J. H. & Ryoo, S. B. Development and Assessment of ADAM3 Ensemble Prediction System. *Sci. Online Lett. Atmos.* **19**, 26–32 (2023).

42. Holben, B. N. *et al.* AERONET—A Federated Instrument Network and Data Archive for Aerosol Characterization. *Remote Sens. Environ.* **66**, 1–16 (1998).

43. Sinyuk, A. *et al.* The AERONET Version 3 aerosol retrieval algorithm, associated uncertainties and comparisons to Version 2. *Atmos. Meas. Tech.* **13**, 3375–3411 (2020).

44. Eck, T. F. *et al.* Wavelength dependence of the optical depth of biomass burning, urban, and desert dust aerosols. *J. Geophys. Res. Atmos.* **104**, 31333–31349 (1999).

45. Schuster, G. L., Dubovik, O. & Holben, B. N. Angstrom exponent and bimodal aerosol size distributions. *J. Geophys. Res. Atmos.* **111**, (2006).

46. Vaswani, A. *et al.* Attention is all you need. *Adv. Neural Inf. Process. Syst.* (2017).

# *Supplementary Information*

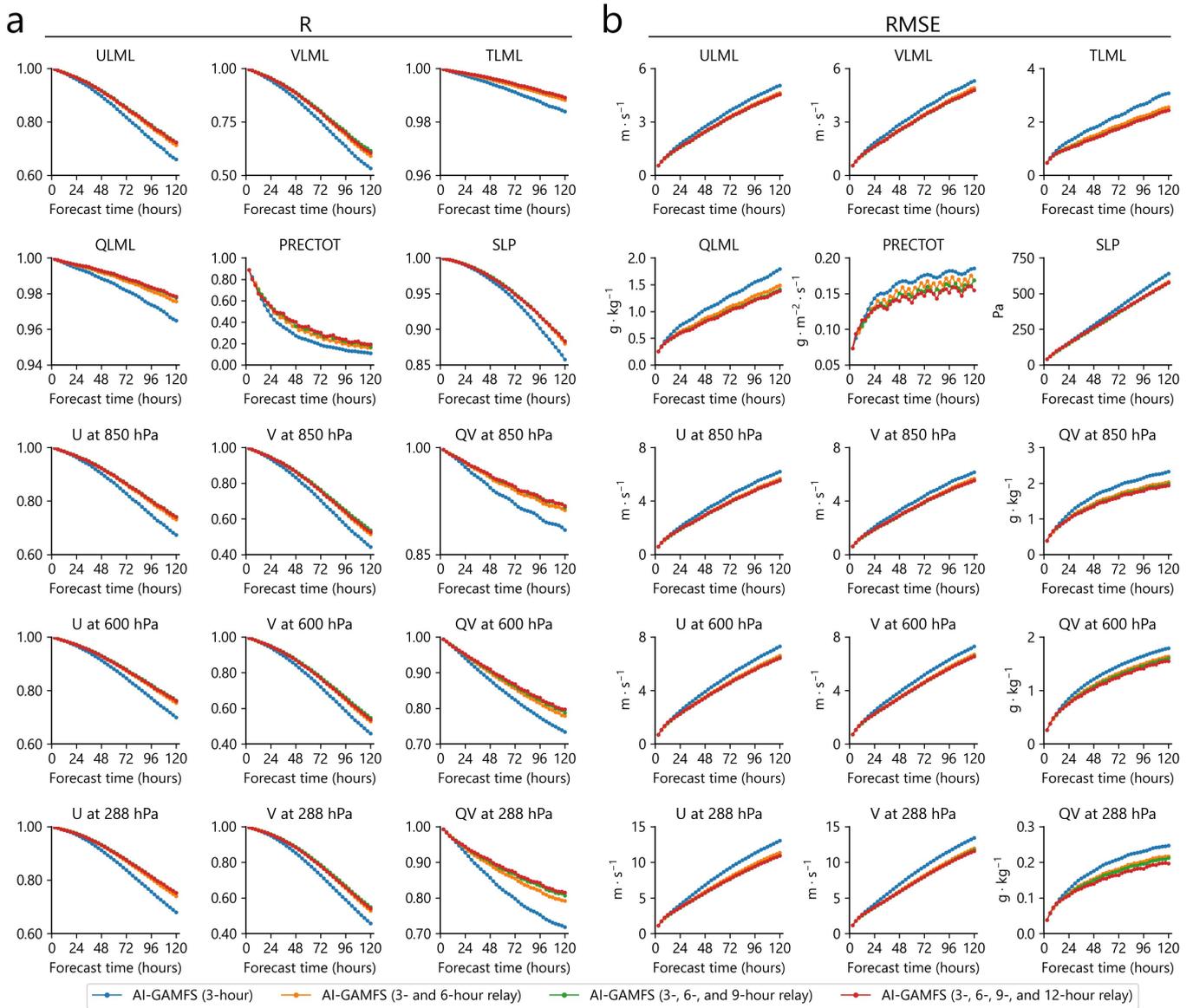

**Figure S1**. **Comparison of meteorological variable forecasting accuracy between single-model and multi-model relay forecasting approaches. a, b,** Accuracy of global 5-day deterministic forecasts for six surface meteorological variables and nine selected upper-level meteorological variables in 2022, showing spatial *R* **(a)** and latitude-weighted RMSE **(b)** over time. The comparisons include the 3-hour single-model, 3- and 6-hour relay model, 3-, 6-, and 9-hour relay model, and 3-, 6-, 9-, and 12-hour relay model. The six surface meteorological variables include surface eastward wind (ULML), surface northward wind (VLML), surface air temperature (TLML), surface specific humidity (QLML), total precipitation (PRECTOT), and sea level pressure (SLP). The nine upper-level meteorological variables include eastward wind (U), northward wind (V), and specific humidity (QV) at 850, 600, and 288 hPa.

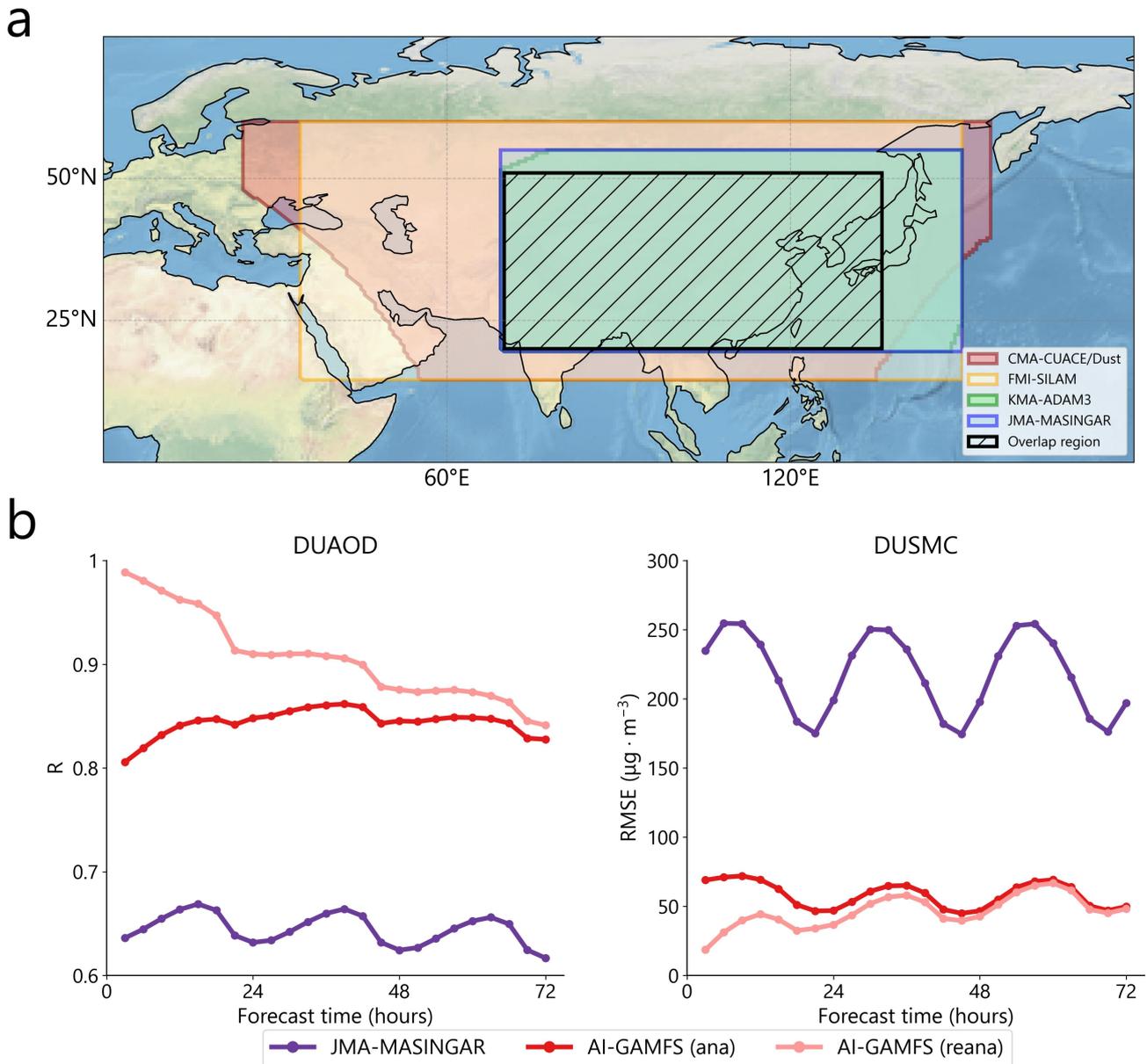

**Figure S2.** **Superior performance of AI-GAMFS in dust forecasting. a,** Spatial distribution of forecast products from four dust forecasting models (i.e., CMA-CUACE, FMI-SILAM, KMA-ADAMS, and JMA-MASINGAR) deployed at the SDS-WAS Asian Regional Centre. The area enclosed by the black borders represents the overlap of the forecast regions from all four models, which is used in this study for comparative analysis and evaluation of the East Asia region. **b,** Comparison of 3-day deterministic forecast accuracy for DUAOD and DUSMC over East Asia in 2023, using MERRA-2 as the reference baseline. The forecast accuracy is evaluated by spatial *R* and latitude-weighted RMSE. The comparison is made between AI-GAMFS (driven by MERRA-2 reanalysis and GEOS-FP analyses) and JMA-MASINGAR. Note that the AI-GAMFS and JMA-MASINGAR forecasts are initialized daily at 10:30 UTC and 12:00 UTC, respectively.

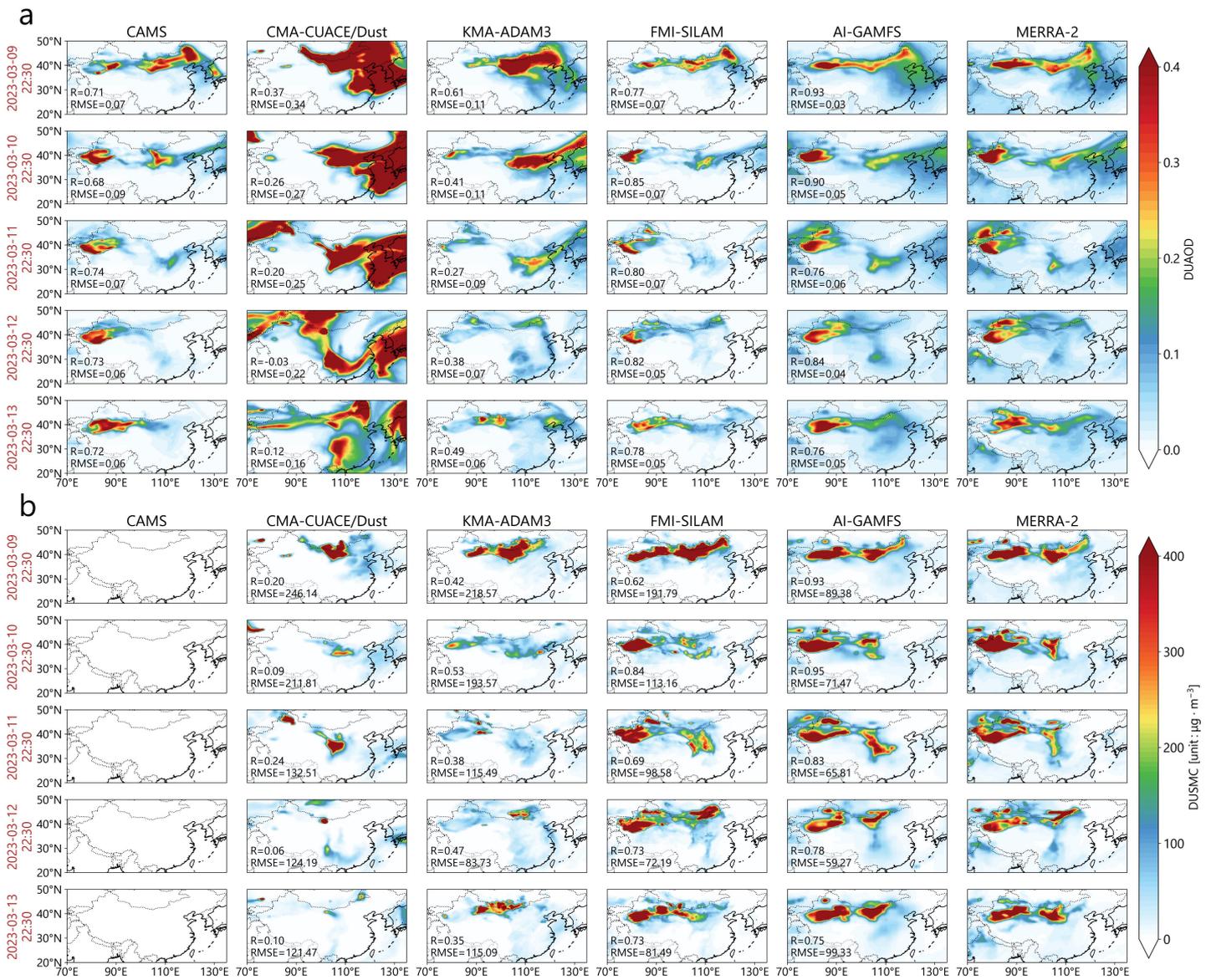

**Figure S3**. **Case study of a mega dust storm event in northern China from March 9 to 13, 2024. a, b,** Forecasts of DUAOD **(a)** and DUSMC **(b)** during the event, produced by AI-GAMFS (driven by GEOS-FP analyses), CAMS, CMA-CUACE/Dust, KMA-ADAMS, and FMI-SILAM models, with a 5-day lead time at 24-hour intervals, compared against MERRA-2 reference data. AI-GAMFS was initialized at 22:30 UTC on March 8, 2024, while the other models were initialized at 00:00 UTC on March 9, 2024. For comparison, the outputs of all models were interpolated to the same spatiotemporal resolution as AI-GAMFS. The overall accuracy metrics (i.e., spatial *R* and latitude-weighted RMSE) for AI-GAMFS and the five dust models relative to MERRA-2 data are shown in the lower-left corner of each panel. Note that DUSMC is not included as a forecast variable for the CAMS model.

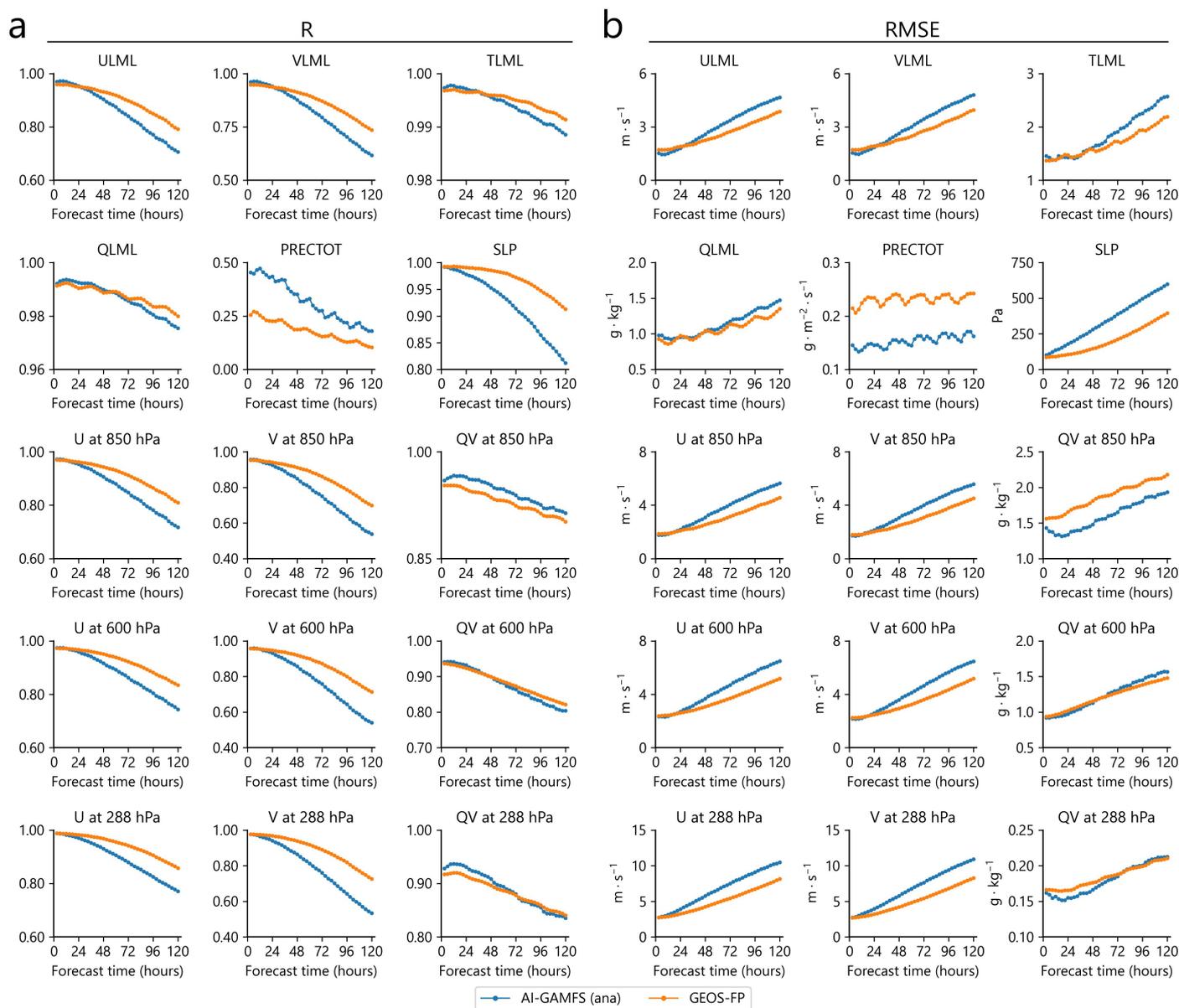

**Figure S4. Comparison of meteorological variable forecasting accuracy between AI-GAMFS and GEOS-FP during operational deployment. a, b,** Comparison of spatial *R* **(a)** and latitude-weighted RMSE **(b)** between AI-GAMFS and GEOS-FP for six surface meteorological variables and nine selected upper-level meteorological variables in 5-day global forecasts with a temporal resolution of 3 hours. MERRA-2 data from July to August 2024 are used as the reference baseline.

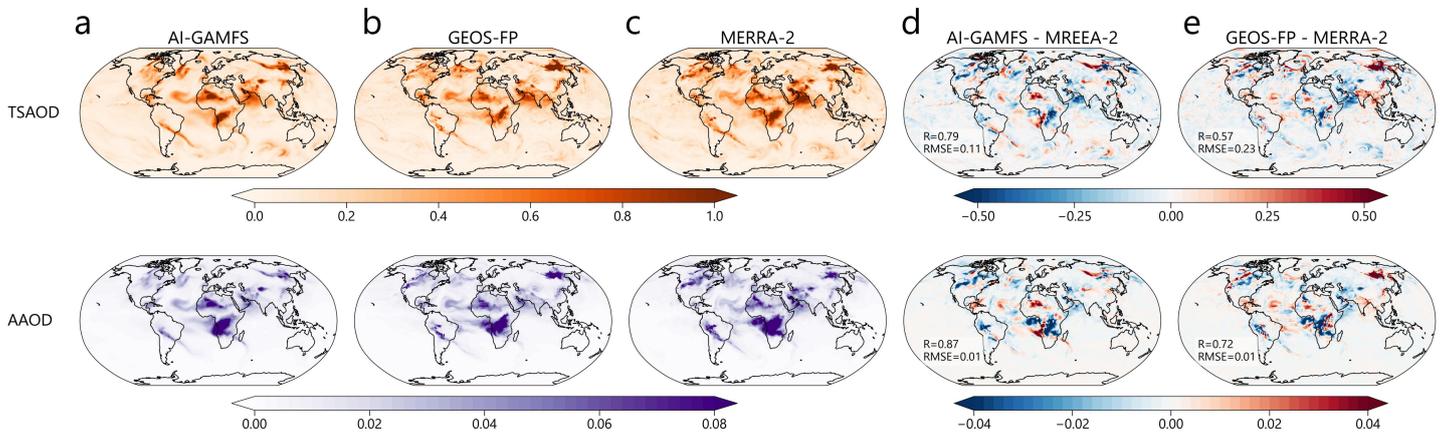

**Figure S5**. **Case study of global aerosol absorption and scattering properties forecasting. a, b, c,** The 3-day lead-time global forecast for TSAOD and AAOD from AI-GAMFS (driven by GEOS-FP analyses) (**a**), GEOS-FP (**b**), and MERRA-2 (**c**), initialized at 22:30 UTC on 26 July 2024. **d, e,** Forecasting errors of AI-GAMFS (**d**) and GEOS-FP (**e**) relative to MERRA-2 reanalysis data. Overall accuracy metrics (i.e., spatial *R* and latitude-weighted RMSE) for AI-GAMFS and GEOS-FP are shown in the lower-left corner of each panel in **d** and **e**.

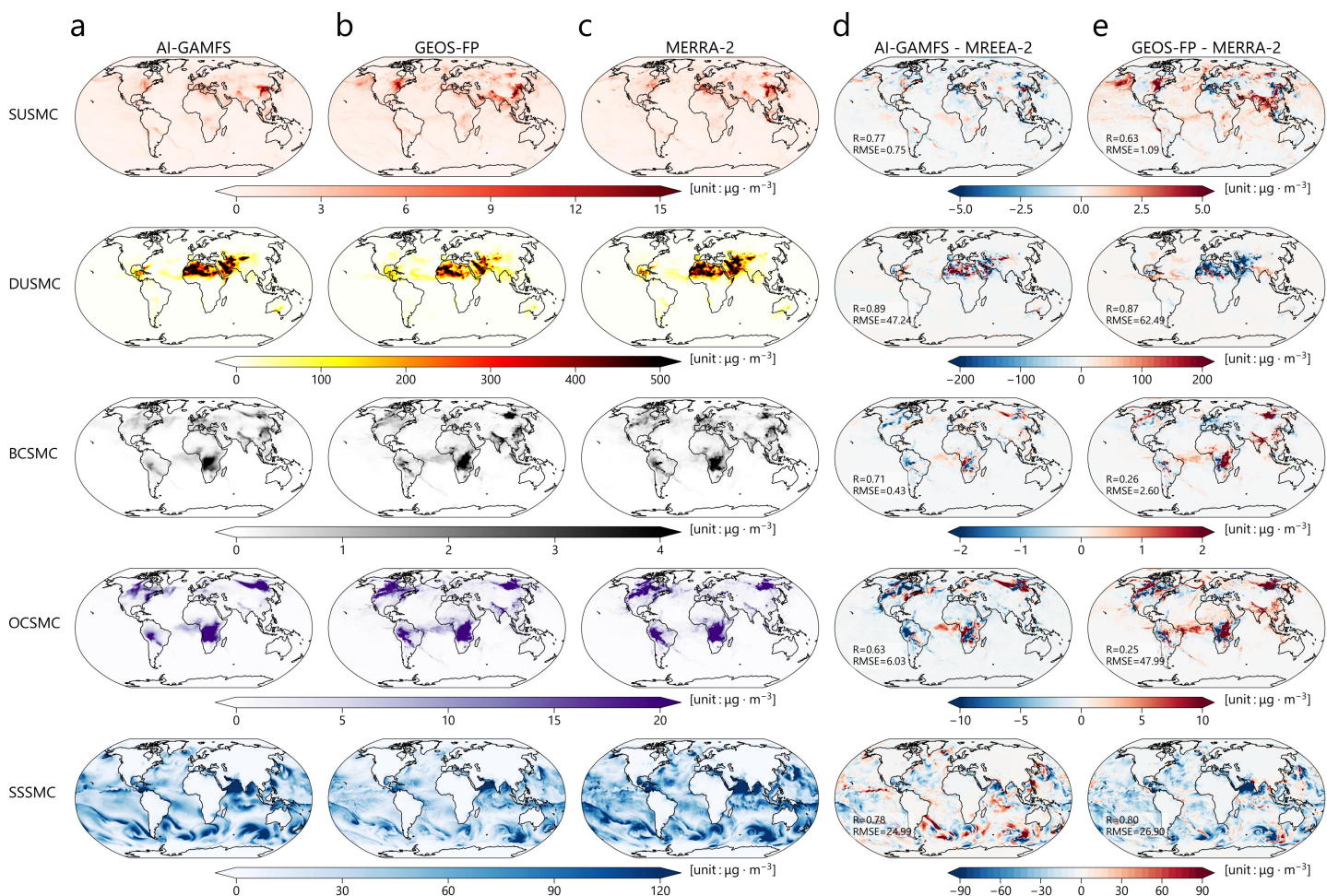

**Figure S6**. **Case study of global aerosol component surface concentrations forecasting. a, b, c,** The 3-day lead-time global forecast for SUSMC, DUSMC, BCSMC, OCSMC, and SSSMC from AI-GAMFS (driven by GEOS-FP analyses) (**a**), GEOS-FP (**b**), and MERRA-2 (**c**), initialized at 22:30 UTC on 26 July 2024. **d, e,** Forecasting errors of AI-GAMFS (**d**) and GEOS-FP (**e**) relative to MERRA-2 reanalysis data. Overall accuracy metrics (i.e., spatial *R* and latitude-weighted RMSE) for AI-GAMFS and GEOS-FP are shown in the lower-left corner of each panel in **d** and **e**.

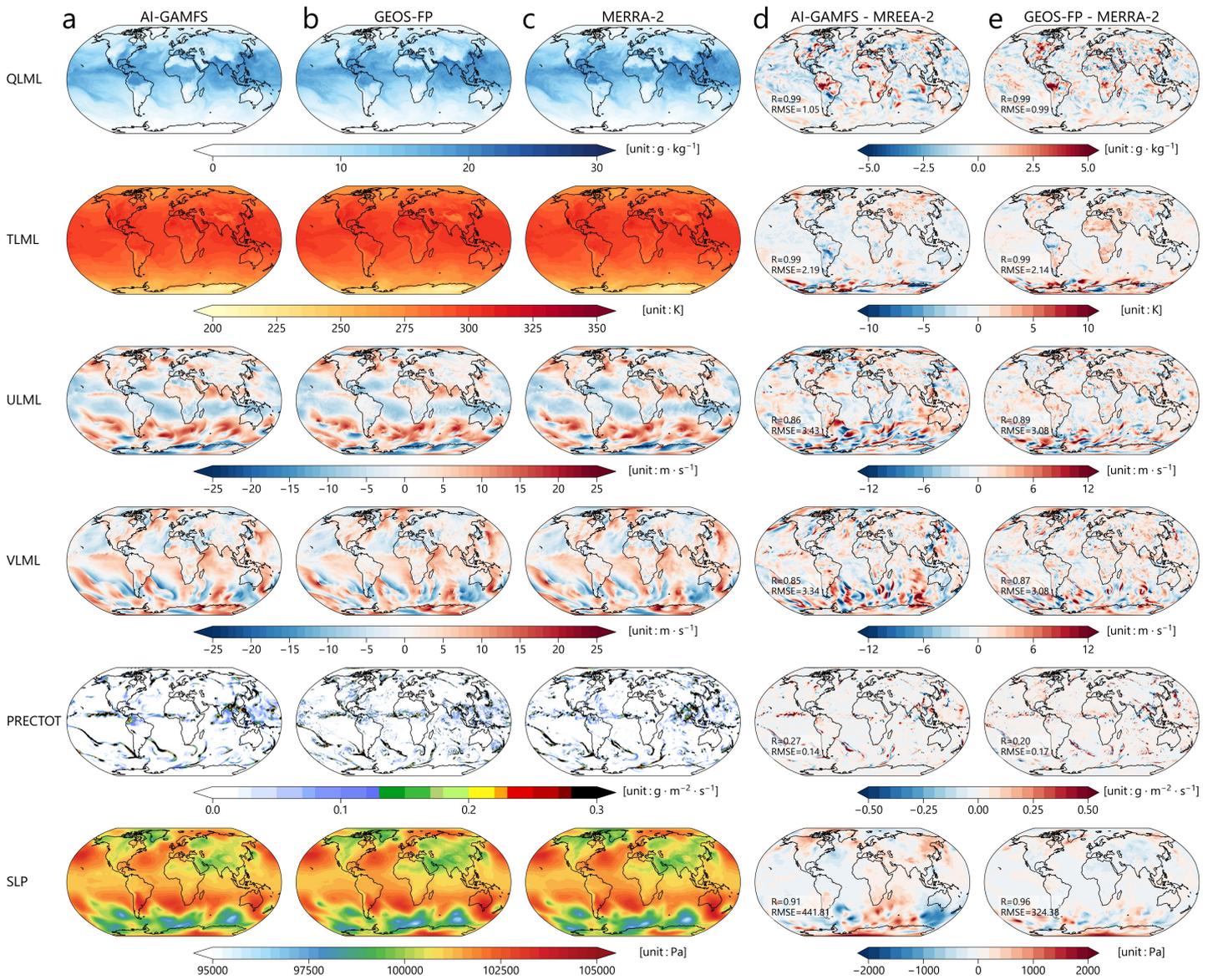

**Figure S7**.**Case study of global surface meteorological forecasts. a, b, c,** The 3-day lead-time global forecast for QLML, TLML, ULML, VLML, PRECTOT, and SLP from AI-GAMFS (driven by GEOS-FP analyses) (**a**), GEOS-FP (**b**), and MERRA-2 (**c**), initialized at 22:30 UTC on 26 July 2024. **d, e,** Forecasting errors of AI-GAMFS (**d**) and GEOS-FP (**e**) relative to MERRA-2 reanalysis data. Overall accuracy metrics (i.e., spatial *R* and latitude-weighted RMSE) for AI-GAMFS and GEOS-FP are shown in the lower-left corner of each panel in **d and e**.

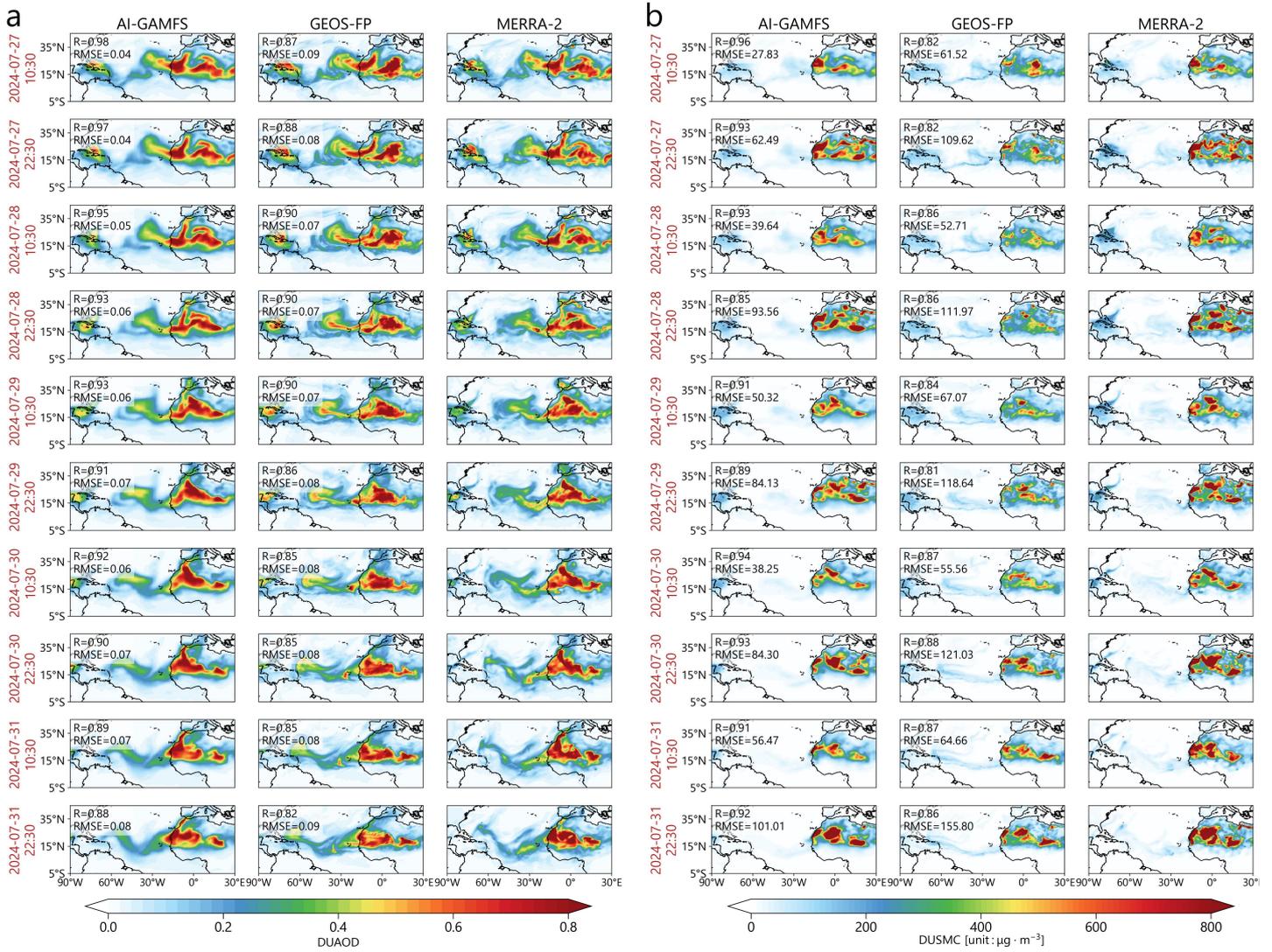

**Figure S8**. **Case study of a trans-Atlantic African dust storm event from July 27 to 31, 2024. a, b,** Forecasts of DUAOD **(a)** and DUSMC **(b)** during the event, produced by AI-GAMFS (driven by GEOS-FP analyses) and GEOS-FP, with a 5-day lead time at 12-hour intervals, compared against MERRA-2 reference data. AI-GAMFS was initialized at 22:30 UTC on July 26, 2024, while GEOS-FP was initialized at 00:00 UTC on July 27, 2024. For comparison, the outputs of GEOS-FP were interpolated to the same spatiotemporal resolution as AI-GAMFS. The overall accuracy metrics (i.e., spatial $R$ and latitude-weighted RMSE) for AI-GAMFS and GEOS-FP relative to MERRA-2 data are shown in the lower-left corner of each panel.

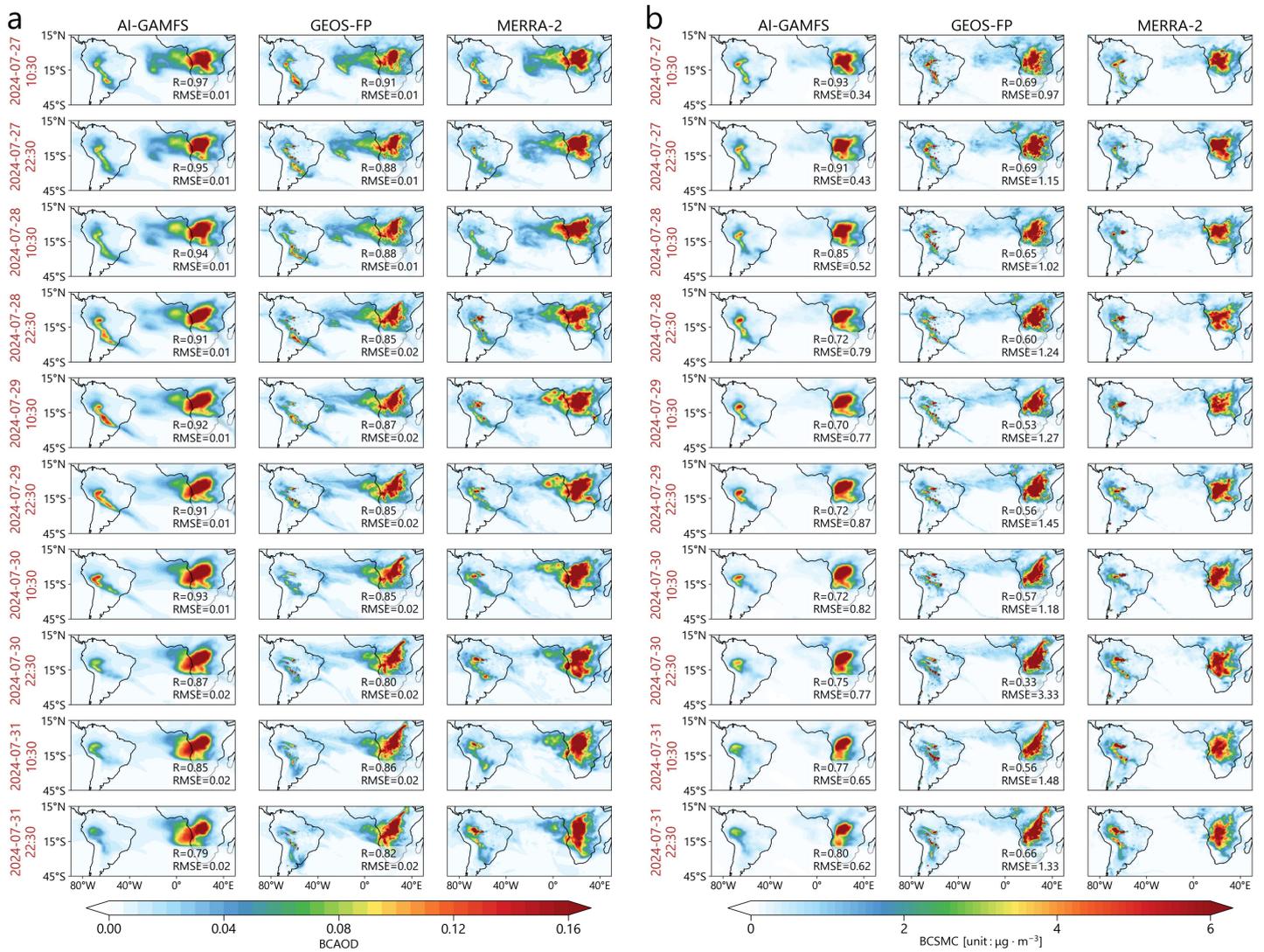

**Figure S9**. Case study of simultaneous biomass burning events in Central Africa and Central South America from July 27 to 31, 2024. **a, b,** Forecasts of BCAOD **(a)** and BCSMC **(b)** during the event, produced by AI-GAMFS (driven by GEOS-FP analyses) and GEOS-FP, with a 5-day lead time at 12-hour intervals, compared against MERRA-2 reference data. AI-GAMFS was initialized at 22:30 UTC on July 26, 2024, while GEOS-FP was initialized at 00:00 UTC on July 27, 2024. For comparison, the outputs of GEOS-FP were interpolated to the same spatiotemporal resolution as AI-GAMFS. The overall accuracy metrics (i.e., spatial *R* and latitude-weighted RMSE) for AI-GAMFS and GEOS-FP relative to MERRA-2 data are shown in the lower-left corner of each panel.

**Table S1. Details of all data used for training, evaluation, validation, and cross-comparison in AI-GAMFS for this study.**

| Data Source | Data type | Collection | Spatial resolution | Temporal resolution | Time range | Variable names (abbrev.)[a] | Pressure levels (hPa)[c] |
|---|---|---|---|---|---|---|---|
| MERRA-2 (reanalysis) | Aerosol | tavg1_2d_aer_Nx | 0.5° × 0.625° | 1-hour | 1980.01–2023.12; 2024.07–2024.08 | AOD, TSAOD, SUAOD, DUAOD, BCAOD, OCAOD, SSAOD, SUSMC, DUSMC, BCSMC, OCSMC, SSSMC | \ |
| | Meteorology | tavg1_2d_flx_Nx | 0.5° × 0.625° | 1-hour | 1980.01–2023.12; 2024.07–2024.08 | QLML, TLML, ULML, VLML, PRECTOT | \ |
| | | tavg3_3d_asm_Nv | 0.5° × 0.625° | 3-hour | 1980.01–2023.12; 2024.07–2024.08 | QV, T, U, SLP[b], V | 985, 925, 850, 800, 700, 600, 525, 413, 288 |
| GEOS-FP (analyses and forecasts) | Aerosol | tavg3_2d_aer_Nx (analyses) | 0.25° × 0.3125° | 3-hour | 2022.01–2023.12; 2024.07–2024.08 | AOD, TSAOD, SUAOD, DUAOD, BCAOD, OCAOD, SSAOD, SUSMC, DUSMC, BCSMC, OCSMC, SSSMC | \ |
| | | tavg3_2d_aer_Nx (forecasts) | 0.25° × 0.3125° | 3-hour | 2024.07–2024.08 | AOD, TSAOD, SUAOD, DUAOD, BCAOD, OCAOD, SSAOD, SUSMC, DUSMC, BCSMC, OCSMC, SSSMC | \ |
| | Meteorology | tavg1_2d_flx_Nx (analyses) | 0.25° × 0.3125° | 1-hour | 2022.01–2023.12; 2024.07–2024.08 | QLML, TLML, ULML, VLML, PRECTOT | \ |
| | | tavg1_2d_flx_Nx (forecasts) | 0.25° × 0.3125° | 1-hour | 2024.07–2024.08 | QLML, TLML, ULML, VLML, PRECTOT | \ |
| | | tavg3_3d_asm_Nv (analyses) | 0.25° × 0.3125° | 3-hour | 2022.01–2023.12; 2024.07–2024.08 | QV, SLP, T, U, V | 985, 925, 850, 800, 700, 600, 525, 413, 288 |
| | | tavg3_3d_asm_Nv (forecasts) | 0.25° × 0.3125° | 3-hour | 2024.07–2024.08 | QV, SLP[b], T, U, V | 985, 925, 850, 800, 700, 600, 525, 413, 288 |
| CAMS (forecasts) | Aerosol | \ | 0.4° × 0.4° | 3-hour | 2023.01–2023.12 | AOD, DUAOD | \ |
| CMA-CUACE/Dust (forecasts) | Aerosol | \ | 0.5° × 0.5° | 3-hour | 2023.01–2023.12 | DUAOD, DUSMC | \ |
| FMI-SILAM (forecasts) | Aerosol | \ | 0.2° × 0.2° | 1-hour | 2023.01–2023.12 | DUAOD, DUSMC | \ |
| JMA-MASINGAR (forecasts) | Aerosol | \ | 0.5° × 0.5° | 1-hour | 2023.01–2023.12 | DUAOD, DUSMC | \ |
| KMA-ADAM3 (forecasts) | Aerosol | \ | 0.5° × 0.5° | 3-hour | 2023.01–2023.12 | DUAOD, DUSMC | \ |
| AERONET | Aerosol | version 3.0, level 2.0 | Point | \ | 2023.01–2023.12 | AOD | \ |

[a] The full names of the 54 variables used in this study are as follows: **Aerosol variables** include total aerosol optical depth (AOD), total scattering AOD (TSAOD), sulfate AOD (SUAOD), dust AOD (DUAOD), black carbon AOD (BCAOD), organic carbon AOD (OCAOD), sea salt AOD (SSAOD), sulfate surface mass concentration (SUSMC), dust surface mass concentration (DUSMC), black carbon surface mass concentration (BCSMC), organic carbon surface mass concentration (OCSMC), and sea salt surface mass concentration (SSSMC). **Surface meteorological variables** include surface specific humidity (QLML), surface air temperature (TLML), surface eastward wind (ULML), surface northward wind (VLML), sea level pressure (SLP), and total precipitation (PRECTOT). **Upper-level meteorological variables** include specific humidity (QV), air temperature (T), eastward wind (U), and northward wind (V).

[b] Note that the SLP variable is stored in *tavg3_3d_asm_Nv*, but it is a single-layer variable.

[c] The nine nominal pressure levels (985, 925, 850, 800, 700, 600, 525, 413, 288 hPa) correspond to the model layers at 72, 68, 63, 60, 56, 53, 51, 48, and 45, respectively.